\documentclass[a4paper,11pt]{article}
\usepackage{jheppub}
\usepackage{subfigure}
\usepackage{graphicx,color,slashed}
\usepackage{xcolor}
\usepackage{tikz}
\usepackage{verbatim}
\usepackage{hyperref}
\usepackage{booktabs}
\usepackage{multirow}
\hypersetup{colorlinks=True, linkcolor=blue,citecolor=green}

\def\be{\begin{eqnarray}}
\def\ed{\end{eqnarray}}
\def\beq{\begin{equation}}
\def\eeq{\end{equation}}
\def\bea{\begin{eqnarray}}
\def\eea{\end{eqnarray}}

\date{\today}

\begin{document}

\title{\bf \Large Dark Transition Magnetic Moments of Majorana Neutrinos Mediated by a Dark Photon}

\author{Haohao Zhang}
\affiliation{School of Physics and State Key Laboratory of Nuclear Physics and Technology, Peking University, Beijing 100871, China}
\emailAdd{2101110115@stu.pku.edu.cn}

\date{\today}

\abstract{Standard Model predictions for Majorana neutrino transition magnetic moments (TMMs) are subject to severe chiral and GIM-like suppressions, rendering them vanishingly small. To dynamically generate a macroscopic TMM, we propose a dark sector framework featuring a $U(1)_D$ gauge symmetry, a vector-like lepton doublet, and two complex dark scalars. We demonstrate that while fermion-radiated loop amplitudes identically cancel due to Majorana self-conjugacy, a chirally enhanced dark TMM is successfully generated exclusively through scalar-radiated loops. This mechanism safely shifts the required chirality flip onto the heavy internal fermion line and utilizes a misaligned double-scalar mixing in flavor space to evade the Majorana antisymmetry prohibition. We systematically confront this tensor portal framework with multi-frontier experimental constraints. Since the dark TMM generation is inextricably linked to charged lepton flavor violation, the internal Yukawa couplings are stringently capped by the latest $\mu \to e \gamma$ limits from MEG II. Concurrently, the visible-dark kinetic mixing portal is heavily bottlenecked by missing energy and mono-photon searches at NA64 and BaBar. Our global phenomenological analysis reveals that the synergistic theoretical upper bound dictated by these indirect high-energy probes completely eclipses the direct scattering constraints from Borexino. This establishes a strict phenomenological hierarchy: high-intensity cLFV probes and accelerator-based dark sector searches jointly possess the overwhelmingly dominant exclusionary power over direct solar neutrino limits for such microscopic magnetic moment models.}

\maketitle
\newpage
\section{Introduction}
\label{sec:intro}

While the Standard Model (SM) strictly assumes neutrinos to be massless---and thus lacking a magnetic moment---a series of pioneering measurements of solar and atmospheric neutrino fluxes~\cite{PhysRevLett.20.1205, PhysRevLett.20.1209, PhysRevLett.63.16, 1991PhRvL..67.3332A, GALLEX:1992gcp} revealed a persistent missing flux anomaly. This long-standing puzzle was ultimately resolved by the definitive observation of neutrino flavor oscillation by the Super-Kamiokande~\cite{PhysRevLett.81.1562} and SNO~\cite{PhysRevLett.87.071301} collaborations. These breakthroughs, further corroborated by subsequent precision terrestrial experiments~\cite{An_2012, An_2016, ye2024multicubickilometreneutrinotelescopewestern}, firmly established that neutrinos possess non-zero masses, providing the first unequivocal evidence of physics beyond the SM ~\cite{Esteban:2020cvm, deSalas:2020pgw} and naturally motivating the study of finite neutrino magnetic moments.
While oscillation data precisely determine the neutrino mass squared differences ($\Delta m_{21}^2$ and $|\Delta m_{31}^2|$), bounds on the absolute mass scale are established by complementary kinematic, neutrinoless double-beta decay, and cosmological probes. Current leading limits restrict the effective electron neutrino mass to $m_{\nu_{e}} < 0.8 \, \mathrm{eV}$ (KATRIN~\cite{KATRIN:2021uub}), the effective Majorana mass to $\langle m_{\beta\beta} \rangle \lesssim 0.036 - 0.156 \, \mathrm{eV}$ (KamLAND-Zen~\cite{Abe_2023, Zhang:2021wjj}), and the cosmological sum of neutrino masses to $\sum m_\nu \lesssim 0.07 \, \mathrm{eV}$ (DESI and Planck~\cite{2020, Adame_2025}). 
A profound theoretical consequence of these non-zero masses is the inevitable generation of a finite neutrino magnetic moment. Unlike charged fermions, neutrinos remain strictly electrically neutral at tree level within the minimally extended Standard Model ($\nu$SM)~\cite{Marciano:1977wx, Bell:2006wi}. Nevertheless, they radiatively acquire a non-vanishing magnetic moment at the one-loop level via interactions with the $W$ boson and charged leptons (see Fig.~\ref{neutromagloop0})~\cite{Giunti_2025}:
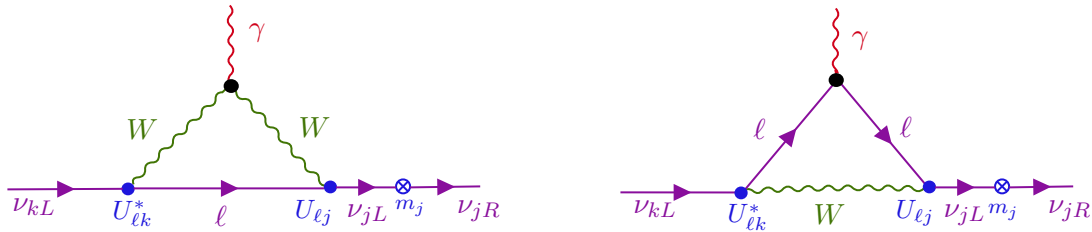
\begin{figure}[!htbp]
    \centering

\tikzset{every picture/.style={line width=0.75pt}} 

\begin{tikzpicture}[x=0.75pt,y=0.75pt,yscale=-1,xscale=1]

\draw [color={rgb, 255:red, 65; green, 117; blue, 5 }  ,draw opacity=1 ]   (406.95,2292.56) .. controls (408.52,2290.57) and (410.25,2290.31) .. (412.16,2291.79) .. controls (413.98,2293.32) and (415.55,2293.2) .. (416.88,2291.44) .. controls (418.65,2289.67) and (420.45,2289.57) .. (422.28,2291.15) .. controls (423.89,2292.74) and (425.46,2292.67) .. (427,2290.94) .. controls (428.62,2289.21) and (430.33,2289.15) .. (432.13,2290.76) .. controls (433.97,2292.37) and (435.6,2292.33) .. (437.03,2290.62) .. controls (438.87,2288.91) and (440.58,2288.86) .. (442.15,2290.49) .. controls (443.73,2292.12) and (445.49,2292.09) .. (447.42,2290.39) .. controls (448.99,2288.7) and (450.58,2288.67) .. (452.21,2290.32) .. controls (453.84,2291.97) and (455.45,2291.95) .. (457.04,2290.26) .. controls (458.64,2288.58) and (460.25,2288.57) .. (461.88,2290.23) .. controls (463.89,2291.88) and (465.7,2291.88) .. (467.29,2290.21) .. controls (468.88,2288.54) and (470.45,2288.55) .. (472.02,2290.22) .. controls (473.94,2291.89) and (475.67,2291.9) .. (477.21,2290.25) .. controls (478.73,2288.6) and (480.39,2288.63) .. (482.2,2290.32) .. controls (483.95,2292.02) and (485.53,2292.05) .. (486.95,2290.41) .. controls (488.67,2288.79) and (490.31,2288.84) .. (491.87,2290.56) .. controls (493.62,2292.29) and (495.39,2292.37) .. (497.17,2290.78) -- (500.29,2290.97) ;
\draw [color={rgb, 255:red, 137; green, 14; blue, 156 }  ,draw opacity=1 ]   (346,2292.5) -- (406.95,2292.56) ;
\draw [shift={(381.48,2292.54)}, rotate = 180.06] [fill={rgb, 255:red, 137; green, 14; blue, 156 }  ,fill opacity=1 ][line width=0.08]  [draw opacity=0] (8.93,-4.29) -- (0,0) -- (8.93,4.29) -- cycle    ;
\draw     ;
\draw [color={rgb, 255:red, 137; green, 14; blue, 156 }  ,draw opacity=1 ]   (406.95,2292.56) -- (454.33,2235.73) ;
\draw [shift={(433.84,2260.3)}, rotate = 129.82] [fill={rgb, 255:red, 137; green, 14; blue, 156 }  ,fill opacity=1 ][line width=0.08]  [draw opacity=0] (8.93,-4.29) -- (0,0) -- (8.93,4.29) -- cycle    ;
\draw [color={rgb, 255:red, 137; green, 14; blue, 156 }  ,draw opacity=1 ]   (500.29,2290.97) -- (454.33,2235.73) ;
\draw [shift={(481.47,2268.35)}, rotate = 230.25] [fill={rgb, 255:red, 137; green, 14; blue, 156 }  ,fill opacity=1 ][line width=0.08]  [draw opacity=0] (8.93,-4.29) -- (0,0) -- (8.93,4.29) -- cycle    ;
\draw [color={rgb, 255:red, 208; green, 2; blue, 27 }  ,draw opacity=1 ]   (151,2198.5) .. controls (152.73,2200.43) and (152.79,2202.12) .. (151.17,2203.58) .. controls (149.56,2205.41) and (149.61,2207.14) .. (151.32,2208.76) .. controls (153.03,2210.28) and (153.07,2211.96) .. (151.45,2213.81) .. controls (149.82,2215.5) and (149.85,2217.08) .. (151.54,2218.56) .. controls (153.23,2220.17) and (153.24,2221.92) .. (151.59,2223.81) .. controls (149.92,2225.36) and (149.91,2226.93) .. (151.55,2228.51) .. controls (153.16,2230.26) and (153.05,2231.92) .. (151.2,2233.49) .. controls (149.8,2235.49) and (150.07,2237.16) .. (152,2238.5) -- (152,2238.5) ;
\draw     ;
\draw [color={rgb, 255:red, 65; green, 117; blue, 5 }  ,draw opacity=1 ]   (103,2288.5) .. controls (102.11,2286.31) and (102.83,2285.13) .. (105.17,2284.95) .. controls (107.6,2284.75) and (108.8,2283.45) .. (108.75,2281.06) .. controls (108.57,2278.83) and (109.73,2277.62) .. (112.22,2277.43) .. controls (114.45,2277.52) and (115.6,2276.34) .. (115.68,2273.87) .. controls (115.64,2271.53) and (116.81,2270.34) .. (119.2,2270.3) .. controls (121.61,2270.23) and (122.76,2269.07) .. (122.64,2266.83) .. controls (122.75,2264.35) and (123.95,2263.14) .. (126.24,2263.21) .. controls (128.54,2263.26) and (129.67,2262.13) .. (129.62,2259.82) .. controls (129.82,2257.27) and (131.09,2256) .. (133.43,2256.02) .. controls (135.77,2256.04) and (136.94,2254.88) .. (136.93,2252.54) .. controls (136.92,2250.19) and (138.08,2249.03) .. (140.43,2249.05) .. controls (142.77,2249.07) and (143.93,2247.91) .. (143.91,2245.58) .. controls (143.88,2243.26) and (145.02,2242.12) .. (147.33,2242.17) -- (150,2239.5) ;
\draw [color={rgb, 255:red, 65; green, 117; blue, 5 }  ,draw opacity=1 ]   (150,2239.5) .. controls (152.37,2239.05) and (153.78,2240.01) .. (154.25,2242.4) .. controls (154.58,2244.77) and (155.87,2245.82) .. (158.12,2245.53) .. controls (160.43,2245.36) and (161.68,2246.53) .. (161.89,2249.02) .. controls (161.87,2251.34) and (163.03,2252.51) .. (165.36,2252.53) .. controls (167.73,2252.62) and (168.88,2253.86) .. (168.79,2256.24) .. controls (168.68,2258.62) and (169.73,2259.82) .. (171.96,2259.83) .. controls (174.36,2260.06) and (175.5,2261.38) .. (175.37,2263.79) .. controls (175.08,2266.01) and (176.14,2267.25) .. (178.55,2267.51) .. controls (180.96,2267.76) and (182.03,2269.01) .. (181.76,2271.25) .. controls (181.67,2273.68) and (182.75,2274.92) .. (185.01,2274.95) .. controls (187.42,2275.14) and (188.61,2276.44) .. (188.57,2278.86) .. controls (188.41,2281.11) and (189.54,2282.28) .. (191.96,2282.37) .. controls (194.36,2282.39) and (195.52,2283.51) .. (195.44,2285.72) .. controls (195.61,2288.13) and (196.9,2289.26) .. (199.3,2289.12) -- (201,2290.5) ;
\draw  [color={rgb, 255:red, 30; green, 22; blue, 220 }  ,draw opacity=1 ][fill={rgb, 255:red, 30; green, 22; blue, 220 }  ,fill opacity=1 ] (97.42,2290.78) .. controls (97.42,2289.27) and (98.67,2288.06) .. (100.21,2288.06) .. controls (101.75,2288.06) and (103,2289.27) .. (103,2290.78) .. controls (103,2292.28) and (101.75,2293.5) .. (100.21,2293.5) .. controls (98.67,2293.5) and (97.42,2292.28) .. (97.42,2290.78) -- cycle ;
\draw  [color={rgb, 255:red, 30; green, 22; blue, 220 }  ,draw opacity=1 ][fill={rgb, 255:red, 30; green, 22; blue, 220 }  ,fill opacity=1 ] (198.42,2289.78) .. controls (198.42,2288.27) and (199.67,2287.06) .. (201.21,2287.06) .. controls (202.75,2287.06) and (204,2288.27) .. (204,2289.78) .. controls (204,2291.28) and (202.75,2292.5) .. (201.21,2292.5) .. controls (199.67,2292.5) and (198.42,2291.28) .. (198.42,2289.78) -- cycle ;
\draw  [color={rgb, 255:red, 30; green, 22; blue, 220 }  ,draw opacity=1 ][fill={rgb, 255:red, 255; green, 255; blue, 255 }  ,fill opacity=1 ] (234,2289.5) .. controls (234,2287.57) and (235.57,2286) .. (237.5,2286) .. controls (239.43,2286) and (241,2287.57) .. (241,2289.5) .. controls (241,2291.43) and (239.43,2293) .. (237.5,2293) .. controls (235.57,2293) and (234,2291.43) .. (234,2289.5) -- cycle ; \draw  [color={rgb, 255:red, 30; green, 22; blue, 220 }  ,draw opacity=1 ] (235.03,2287.03) -- (239.97,2291.97) ; \draw  [color={rgb, 255:red, 30; green, 22; blue, 220 }  ,draw opacity=1 ] (239.97,2287.03) -- (235.03,2291.97) ;
\draw [color={rgb, 255:red, 137; green, 14; blue, 156 }  ,draw opacity=1 ]   (40,2291) -- (97.54,2290.72) ;
\draw [shift={(73.77,2290.84)}, rotate = 179.73] [fill={rgb, 255:red, 137; green, 14; blue, 156 }  ,fill opacity=1 ][line width=0.08]  [draw opacity=0] (8.93,-4.29) -- (0,0) -- (8.93,4.29) -- cycle    ;
\draw [color={rgb, 255:red, 137; green, 14; blue, 156 }  ,draw opacity=1 ]   (103,2290.5) -- (198,2290.5) ;
\draw [shift={(155.5,2290.5)}, rotate = 180] [fill={rgb, 255:red, 137; green, 14; blue, 156 }  ,fill opacity=1 ][line width=0.08]  [draw opacity=0] (8.93,-4.29) -- (0,0) -- (8.93,4.29) -- cycle    ;
\draw [color={rgb, 255:red, 137; green, 14; blue, 156 }  ,draw opacity=1 ]   (204,2290) -- (234,2290) ;
\draw [shift={(224,2290)}, rotate = 180] [fill={rgb, 255:red, 137; green, 14; blue, 156 }  ,fill opacity=1 ][line width=0.08]  [draw opacity=0] (8.93,-4.29) -- (0,0) -- (8.93,4.29) -- cycle    ;
\draw [color={rgb, 255:red, 137; green, 14; blue, 156 }  ,draw opacity=1 ]   (241,2290) -- (276,2289.5) ;
\draw [shift={(263.5,2289.68)}, rotate = 179.18] [fill={rgb, 255:red, 137; green, 14; blue, 156 }  ,fill opacity=1 ][line width=0.08]  [draw opacity=0] (8.93,-4.29) -- (0,0) -- (8.93,4.29) -- cycle    ;
\draw  [color={rgb, 255:red, 0; green, 0; blue, 0 }  ,draw opacity=1 ][fill={rgb, 255:red, 0; green, 0; blue, 0 }  ,fill opacity=1 ] (148.42,2239.03) .. controls (148.42,2237.39) and (149.89,2236.06) .. (151.71,2236.06) .. controls (153.53,2236.06) and (155,2237.39) .. (155,2239.03) .. controls (155,2240.67) and (153.53,2242) .. (151.71,2242) .. controls (149.89,2242) and (148.42,2240.67) .. (148.42,2239.03) -- cycle ;
\draw [color={rgb, 255:red, 208; green, 2; blue, 27 }  ,draw opacity=1 ]   (453.5,2200) .. controls (455.17,2202.09) and (455.16,2203.88) .. (453.49,2205.36) .. controls (451.82,2207.21) and (451.82,2208.8) .. (453.48,2210.11) .. controls (455.13,2211.76) and (455.12,2213.53) .. (453.44,2215.4) .. controls (451.75,2217.09) and (451.73,2218.71) .. (453.38,2220.26) .. controls (455.02,2221.92) and (454.99,2223.55) .. (453.29,2225.14) .. controls (451.58,2226.77) and (451.53,2228.43) .. (453.13,2230.11) .. controls (454.65,2231.99) and (454.75,2232.54) .. (453.42,2231.76) -- (454,2234.5) ;
\draw  [color={rgb, 255:red, 0; green, 0; blue, 0 }  ,draw opacity=1 ][fill={rgb, 255:red, 0; green, 0; blue, 0 }  ,fill opacity=1 ] (450.42,2236.03) .. controls (450.42,2234.39) and (451.89,2233.06) .. (453.71,2233.06) .. controls (455.53,2233.06) and (457,2234.39) .. (457,2236.03) .. controls (457,2237.67) and (455.53,2239) .. (453.71,2239) .. controls (451.89,2239) and (450.42,2237.67) .. (450.42,2236.03) -- cycle ;
\draw  [color={rgb, 255:red, 30; green, 22; blue, 220 }  ,draw opacity=1 ][fill={rgb, 255:red, 30; green, 22; blue, 220 }  ,fill opacity=1 ] (403.42,2292.78) .. controls (403.42,2291.27) and (404.67,2290.06) .. (406.21,2290.06) .. controls (407.75,2290.06) and (409,2291.27) .. (409,2292.78) .. controls (409,2294.28) and (407.75,2295.5) .. (406.21,2295.5) .. controls (404.67,2295.5) and (403.42,2294.28) .. (403.42,2292.78) -- cycle ;
\draw  [color={rgb, 255:red, 30; green, 22; blue, 220 }  ,draw opacity=1 ][fill={rgb, 255:red, 30; green, 22; blue, 220 }  ,fill opacity=1 ] (497.42,2289.78) .. controls (497.42,2288.27) and (498.67,2287.06) .. (500.21,2287.06) .. controls (501.75,2287.06) and (503,2288.27) .. (503,2289.78) .. controls (503,2291.28) and (501.75,2292.5) .. (500.21,2292.5) .. controls (498.67,2292.5) and (497.42,2291.28) .. (497.42,2289.78) -- cycle ;
\draw  [color={rgb, 255:red, 30; green, 22; blue, 220 }  ,draw opacity=1 ][fill={rgb, 255:red, 255; green, 255; blue, 255 }  ,fill opacity=1 ] (533,2289.5) .. controls (533,2287.57) and (534.57,2286) .. (536.5,2286) .. controls (538.43,2286) and (540,2287.57) .. (540,2289.5) .. controls (540,2291.43) and (538.43,2293) .. (536.5,2293) .. controls (534.57,2293) and (533,2291.43) .. (533,2289.5) -- cycle ; \draw  [color={rgb, 255:red, 30; green, 22; blue, 220 }  ,draw opacity=1 ] (534.03,2287.03) -- (538.97,2291.97) ; \draw  [color={rgb, 255:red, 30; green, 22; blue, 220 }  ,draw opacity=1 ] (538.97,2287.03) -- (534.03,2291.97) ;
\draw [color={rgb, 255:red, 137; green, 14; blue, 156 }  ,draw opacity=1 ]   (503,2290) -- (533,2290) ;
\draw [shift={(523,2290)}, rotate = 180] [fill={rgb, 255:red, 137; green, 14; blue, 156 }  ,fill opacity=1 ][line width=0.08]  [draw opacity=0] (8.93,-4.29) -- (0,0) -- (8.93,4.29) -- cycle    ;
\draw [color={rgb, 255:red, 137; green, 14; blue, 156 }  ,draw opacity=1 ]   (540,2290) -- (572,2289.5) ;
\draw [shift={(561,2289.67)}, rotate = 179.1] [fill={rgb, 255:red, 137; green, 14; blue, 156 }  ,fill opacity=1 ][line width=0.08]  [draw opacity=0] (8.93,-4.29) -- (0,0) -- (8.93,4.29) -- cycle    ;

\draw (41,2294) node [anchor=north west][inner sep=0.75pt]  [color={rgb, 255:red, 137; green, 14; blue, 156 }  ,opacity=1 ]  {$\nu _{kL}$};
\draw (142,2298) node [anchor=north west][inner sep=0.75pt]  [color={rgb, 255:red, 137; green, 14; blue, 156 }  ,opacity=1 ]  {$\ell $};
\draw (208,2297) node [anchor=north west][inner sep=0.75pt]  [font=\normalsize,color={rgb, 255:red, 137; green, 14; blue, 156 }  ,opacity=1 ]  {$\nu _{jL}$};
\draw (98,2254) node [anchor=north west][inner sep=0.75pt]  [color={rgb, 255:red, 65; green, 117; blue, 5 }  ,opacity=1 ]  {$W$};
\draw (184,2253) node [anchor=north west][inner sep=0.75pt]  [color={rgb, 255:red, 65; green, 117; blue, 5 }  ,opacity=1 ]  {$W$};
\draw (159,2206) node [anchor=north west][inner sep=0.75pt]  [color={rgb, 255:red, 208; green, 2; blue, 27 }  ,opacity=1 ]  {$\gamma $};
\draw (91,2295) node [anchor=north west][inner sep=0.75pt]  [font=\small,color={rgb, 255:red, 30; green, 22; blue, 220 }  ,opacity=1 ]  {$U_{\ell k}^{*}$};
\draw (182,2295) node [anchor=north west][inner sep=0.75pt]  [font=\small,color={rgb, 255:red, 30; green, 22; blue, 220 }  ,opacity=1 ]  {$U_{\ell j}$};
\draw (263,2295) node [anchor=north west][inner sep=0.75pt]  [font=\normalsize,color={rgb, 255:red, 137; green, 14; blue, 156 }  ,opacity=1 ]  {$\nu _{jR}$};
\draw (232,2295) node [anchor=north west][inner sep=0.75pt]  [font=\scriptsize,color={rgb, 255:red, 30; green, 22; blue, 220 }  ,opacity=1 ]  {$m_{j}$};
\draw (351,2295) node [anchor=north west][inner sep=0.75pt]  [color={rgb, 255:red, 137; green, 14; blue, 156 }  ,opacity=1 ]  {$\nu _{kL}$};
\draw (482,2295) node [anchor=north west][inner sep=0.75pt]  [font=\small,color={rgb, 255:red, 30; green, 22; blue, 220 }  ,opacity=1 ]  {$U_{\ell j}$};
\draw (398,2297) node [anchor=north west][inner sep=0.75pt]  [font=\small,color={rgb, 255:red, 30; green, 22; blue, 220 }  ,opacity=1 ]  {$U_{\ell k}^{*}$};
\draw (506,2297) node [anchor=north west][inner sep=0.75pt]  [font=\normalsize,color={rgb, 255:red, 137; green, 14; blue, 156 }  ,opacity=1 ]  {$\nu _{jL}$};
\draw (558,2295) node [anchor=north west][inner sep=0.75pt]  [font=\normalsize,color={rgb, 255:red, 137; green, 14; blue, 156 }  ,opacity=1 ]  {$\nu _{jR}$};
\draw (528,2297) node [anchor=north west][inner sep=0.75pt]  [font=\scriptsize,color={rgb, 255:red, 30; green, 22; blue, 220 }  ,opacity=1 ]  {$m_{j}$};
\draw (460,2210) node [anchor=north west][inner sep=0.75pt]  [color={rgb, 255:red, 208; green, 2; blue, 27 }  ,opacity=1 ]  {$\gamma $};
\draw (441,2297) node [anchor=north west][inner sep=0.75pt]  [color={rgb, 255:red, 65; green, 117; blue, 5 }  ,opacity=1 ]  {$W$};
\draw (411,2254) node [anchor=north west][inner sep=0.75pt]  [color={rgb, 255:red, 137; green, 14; blue, 156 }  ,opacity=1 ]  {$\ell $};
\draw (484,2253) node [anchor=north west][inner sep=0.75pt]  [color={rgb, 255:red, 137; green, 14; blue, 156 }  ,opacity=1 ]  {$\ell $};

\end{tikzpicture}
    \caption{Feynman Diagram of Neutrino Magnetic Moment.}
    \label{neutromagloop0}
\end{figure}

It is important to note that the specific expression for the magnetic moment depends on whether the neutrino is a Dirac or a Majorana fermion. 
If the neutrino is a Dirac fermion, according to calculations by Lee, Shrock, and Fujikawa et al. \cite{Lee:1977tib, Fujikawa:1980yx}, the expression for the diagonal magnetic moment is:
\begin{equation}\label{eq:mag1}
    \mu_{\nu}^{ii} = \frac{3eG_{F}m_{\nu_i}}{8\sqrt{2}\pi^{2}} \approx 3.2\times10^{-19}\left(\frac{m_{\nu_i}}{1\text{eV}}\right)\mu_{B},
\end{equation}
where $G_F$ is the Fermi constant, and $\mu_B = \frac{e}{2m_e}$ is the Bohr magneton. 
Conversely, if it is a Majorana fermion, the CPT theorem requires that $\mu_{\text{particle}} = -\mu_{\text{antiparticle}}$, meaning the diagonal magnetic moment is strictly zero. Naturally, one can also extract the bilinear part $B_{ij} = \bar{\nu}_i \sigma^{\alpha\beta} \nu_j$ from $\mathcal{L}_{mag} = \frac{1}{2} \mu_{ij} \bar{\nu}_i^{c} \sigma^{\alpha\beta} \nu_j F_{\alpha\beta}$. By utilizing the anticommutation property of fermions and the Majorana condition $\nu = \nu^c = C \bar{\nu}^T$, one can also arrive at the conclusion that the diagonal magnetic moment is zero. Thus, in this case, neutrinos only possess transition magnetic moments (TMM) $\mu_{\nu}^{ij} (i \neq j)$ connecting different mass eigenstates. This distinction is one of the primary motivations for our study of the dark transition magnetic moment in this thesis.
Unlike the Dirac case, the situation for Majorana neutrinos within the Standard Model is far more restrictive. Early research by Shrock et al. \cite{Shrock:1982sc} pointed out that the Majorana transition magnetic moment not only requires an internal chiral flip via $m_\nu$, but is also suppressed by a GIM-like mechanism due to the unitarity of the PMNS matrix. Its magnetic moment contribution must rely on the mass differences of internal charged leptons, thereby introducing an additional dimensionless suppression factor $(m_\ell^2/M_W^2)$:
\begin{equation}
\mu_{ij}^{\text{M}} \propto e G_F (m_i + m_j) \sum_{\ell} U_{i\ell} U^{*}_{j\ell} \frac{m_{\ell}^2}{M_W^2}
\end{equation}
Consequently, the Standard Model prediction for the Majorana transition magnetic moment is vanishingly small, $\sim \mathcal{O}(10^{-23}) \mu_B$, residing far below any foreseeable experimental sensitivity. Experimentally, terrestrial bounds on the neutrino magnetic moment are primarily derived from neutrino-electron elastic scattering. In this process, the photon-exchange contribution exhibits an infrared divergence proportional to $1/T_e$ (where $T_e$ is the electron recoil energy), dominating over the flat weak-interaction background at low energies. Exploiting this low-threshold enhancement, reactor and solar neutrino experiments such as GEMMA~\cite{Beda:2012zz} and Borexino~\cite{Agostini_2017} have constrained the effective magnetic moment to $\mu_\nu < 2.9 \times 10^{-11}\mu_B$ and $2.8 \times 10^{-11} \mu_B$, respectively. More recently, next-generation liquid xenon dark matter detectors, including XENONnT~\cite{PhysRevLett.129.161805} and LUX-ZEPLIN~\cite{PhysRevLett.131.041002}, have leveraged their ultra-low backgrounds to further push these limits down to $\mathcal{O}(10^{-12})\mu_B$~\cite{XENON:2022ltv, LZ:2022lsv, Beda:2012zz}. This reveals a staggering theoretical-experimental gap of at least nine orders of magnitude. Any future observation of a neutrino magnetic moment near the current experimental reach would unambiguously sever the rigid proportionality between the magnetic moment and neutrino mass, providing a smoking-gun signature for new physics. Exploring novel theoretical mechanisms that decouple these two fundamental properties---such as the dark transition magnetic moment proposed in this work---is therefore highly motivated to bridge this vast divide. 

To resolve this theoretical-experimental gap, extending the SM with a secluded dark sector provides a natural framework. The dark photon ($A'$), arising from an additional $U(1)_D$ gauge symmetry, serves as a well-established portal connecting the visible and dark sectors~\cite{Holdom:1985ag, Essig:2013lka, Fabbrichesi:2020wbt, Caputo:2021eaa, Agrawal:2021dbo, Antel:2023hkf}. Recent phenomenological efforts frequently employ such dark photon scenarios to address diverse anomalies, ranging from light dark matter interactions to non-standard neutrino properties (e.g., the neutrino dipole portal)~\cite{Giunti:2014ixa, Magill:2018jla,  Brdar:2020quo, Plestid:2020ssy, Abdullahi:2022jlv, Brdar:2023tmi, Huang:2022pce}. By coupling to this dark gauge field rather than the SM photon, neutrinos can acquire a macroscopic dark transition magnetic moment, effectively evading the stringent structural suppressions inherent in the SM.

Dynamically realizing this dark transition magnetic moment requires a mediator that facilitates both the dark sector connection and the necessary chirality flip. Our framework achieves this through Vector-Like Lepton doublets and complex dark scalars. VLLs are prevalent in BSM model-building, often invoked to explain charged lepton flavor violation, the enduring muon $(g-2)$ anomaly~\cite{Poh:2017tfo, Muong-2:2023cdq}, and radiative neutrino mass generation~\cite{Kannike:2011ng, Falkowski:2013jya, Dermisek:2013gta, Crivellin:2018qmi, Kawamura:2022muq, Altmannshofer:2023hkn}. Moreover, these heavy fermions remain a primary target for direct LHC searches in recent Run 3 analyses~\cite{Bhattiprolu:2019vdu, CMS:2019hsm, Cao:2023smj}. Intertwining VLLs with a misaligned multi-scalar mechanism leads to a chirally-enhanced dark transition magnetic moment, directly subjecting the theory to synergistic constraints from high-intensity flavor experiments (e.g., the latest MEG II results~\cite{MEGII:2023ltw}) and active dark-sector probes (e.g., NA64 and Belle II~\cite{NA64:2023wbi, Belle-II:2022yaw}).

The remainder of this paper is organized as follows. Section \ref{sec:framework} details the construction of the dual-scalar framework, introducing the model Lagrangian and the flavor-space misalignment mechanism, followed by a complete one-loop analytical derivation of the dark transition magnetic moment. In Section \ref{sec:global_constraints}, we conduct a comprehensive phenomenological analysis. We systematically evaluate the severe visible-sector bounds imposed by $\mu \to e \gamma$ alongside the dark-sector constraints from accelerator searches, culminating in a joint parameter space analysis confronted by Borexino direct detection limits. Finally, Section \ref{sec:conclusion} summarizes our core findings and conclusions.

\section{The Dual-Scalar Framework}
\label{sec:framework}

\subsection{Dark Sector Lagrangian}
Without loss of generality, we restrict our analysis to the active electron and muon generations to investigate lepton flavor violation phenomena. The model extends the Standard Model gauge symmetry by a dark $U(1)_D$ group and introduces a vector-like fermion doublet $\Psi = (N, E)^T$ alongside two SM-singlet complex dark scalars, $S_1$ and $S_2$. The corresponding gauge representations and quantum number assignments under $SU(2)_L \times U(1)_Y \times U(1)_D$ are summarized in Table~\ref{tab:quantum_numbers}.

\begin{table}[!htbp]
    \centering 
    \renewcommand{\arraystretch}{1.3}
    \begin{tabular}{lcccc}
        \toprule
         \textbf{Fields} & \textbf{$SU(2)_L$} & \textbf{$U(1)_Y$} & \textbf{$U(1)_D$} & \textbf{$U(1)_{EM}$} \\
        \midrule
        $L_{\alpha L}$ & $\mathbf{2}$ & $-1/2$ & $0$ & $0/-1$ \\
        $\Psi = (N, E)^T$ & $\mathbf{2}$ & $-1/2$ & $+1$ & $0/-1$ \\
         $S_1$ & $\mathbf{1}$ & $0$ & $-1$ & $0$  \\
         $S_2$ & $\mathbf{1}$ & $0$ & $-1$ & $0$  \\
        \bottomrule
    \end{tabular}
    \caption{Gauge group representations and quantum number assignments for relevant fermion and scalar fields in the model.}
    \label{tab:quantum_numbers}
\end{table}

By construction, all SM fields are explicitly uncharged under $U(1)_D$, ensuring that the dark photon $A'_\mu$ couples exclusively to $\Psi$ and $S_a$ at tree level. Furthermore, the gauge invariance of the effective Yukawa operator $\overline{L_{\alpha L}} \Psi_R S_a$ uniquely fixes the electric charge of the lower doublet component $E$ to $-1$. The complete Lagrangian governing this model is thus composed of the SM terms, the kinetic and mass terms for the new dark fields, the portal Yukawa interactions, and a lepton-number-violating Majorana mass insertion:
\begin{align}
    \mathcal{L} &= \mathcal{L}_{\text{SM}} + \mathcal{L}_{\text{new}} + \mathcal{L}_{\text{int}} + \mathcal{L}_{\text{LNV}}, \\
    \mathcal{L}_{\text{new}} &= \bar{\Psi} i \slashed{D} \Psi - M_N \overline{N} N - M_E \overline{E} E + \sum_{a=1,2} \left[ (D_\mu S_a)^\dagger (D^\mu S_a) - M_{S_a}^2 S_a^\dagger S_a \right], \\
    \mathcal{L}_{\text{int}} &= - \sum_{\alpha = e, \mu} \sum_{a=1,2} y_{\alpha a} \bar{L}_{\alpha} P_R \Psi S_a + \text{h.c.} \nonumber \\
    &= - \sum_{\alpha = e, \mu} \sum_{a=1,2} y_{\alpha a} \left( \bar{\nu}_{\alpha L} N_R + \bar{\ell}_{\alpha L} E_R \right) S_a + \text{h.c.}, \label{eq:neuyua} \\
    \mathcal{L}_{\text{LNV}} &= -\frac{1}{2} m \left( \overline{N_L^c} N_L + \overline{N_R^c} N_R \right) + \text{h.c.} , \label{eq:majorana_insertion}
\end{align}
Here, $\mathcal{L}_{\text{new}}$ describes the gauge interaction kinetic terms of the vector-like fermion doublet $\Psi$, as well as the independent Dirac mass terms $M_N$ and $M_E$ acquired by the neutral component $N$ and the charged component $E$, respectively, after electroweak symmetry breaking.

At low energies, integrating out the heavy vector-like fermions and dark scalars yields effective interactions between the light Majorana neutrinos and the dark photon. From an effective field theory perspective, gauge symmetry and Lorentz invariance generally restrict $\nu$-$A'$ couplings to scalar, vector, and tensor structures. However, for Majorana fermions, a diagonal vector coupling ($\bar{\nu}_i \gamma^\mu \nu_i A'_\mu$) identically vanishes due to self-conjugacy, while off-diagonal vector couplings explicitly violate the gauge invariance of the Abelian $U(1)_D$ group. Although scalar couplings ($\bar{\nu}^c \nu \phi$) remain theoretically viable and offer rich cosmological phenomenology---such as connections to Majoron models or resolutions to the Hubble tension via secret interactions~\cite{Chikashige:1980ui, Gelmini:1980re, Escudero_2020, farzan2018neutrinooscillationsnonstandardinteractions, Berryman_2023}---this work focuses exclusively on the tensor portal. Specifically, our model dynamically generates a dark magnetic dipole operator:
\begin{equation}\label{eq:dark_mag}
    \mathcal{L}_{\text{eff}} \supset \frac{1}{2} \mu_{ij}^{D} \bar{\nu}_i^c \sigma_{\mu\nu} \nu_j F'^{\mu\nu} + \text{h.c.}
\end{equation}
The antisymmetric nature of the fermion bilinear under index exchange enforces $\mu_{ii}^D = 0$. Consequently, this interaction strictly induces off-diagonal dark transition magnetic moments, inherently carrying lepton flavor violation and providing a clean, background-free signature for new physics~\cite{Magill_2018,Essig:2013lka, Alexander:2016aln, Battaglieri:2017aum}.

To bridge this dark sector with Standard Model observables, we incorporate the ubiquitous gauge kinetic mixing between the $U(1)_D$ and the SM hypercharge~\cite{Okun:1982xi, Galison:1983pa, Holdom:1985ag}. The relevant terms in the low-energy effective Lagrangian are parameterized as:
\begin{equation}
    \mathcal{L}_{\text{mix}} \supset -\frac{1}{4} F_{\mu\nu} F^{\mu\nu} - \frac{1}{4} F'_{\mu\nu} F'^{\mu\nu} + \frac{\epsilon}{2} F_{\mu\nu} F'^{\mu\nu} + \frac{1}{2} m_{A'}^2 A_\mu A'^\mu,
\end{equation}
where $\epsilon \ll 1$ is the kinetic mixing parameter. Redefining the gauge fields to canonicalize the kinetic terms ($A_\mu \to A_\mu + \epsilon X_\mu$) induces a direct coupling between the dark photon and the SM electromagnetic current, $\mathcal{L}_{\text{int}} \supset \epsilon e A'_{\mu} J_{\text{em}}^\mu$. We safely neglect $Z-A'$ mass mixing, as its effects are severely suppressed by a $(q^2/m_Z^2)$ factor at the sub-GeV energy scales relevant to fixed-target and neutrino scattering experiments.

\subsection{Mass Basis and Misalignment}

To introduce the flavor-space misalignment requisite for evading the exact cancellation of the Majorana transition magnetic moment, we incorporate two generations of complex dark scalars $S_a$ ($a=1, 2$). The pertinent Yukawa interactions in the flavor basis read:
\begin{equation}
    \mathcal{L}_{\nu N S} = - \sum_{\alpha=e, \mu} \sum_{a=1,2} y_{\alpha a} \bar{\nu}_{\alpha L} N_R S_a + \text{h.c.}
\end{equation}

Rotating the active neutrinos into the three physical mass eigenstates via the PMNS matrix ($\nu_{\alpha L} = \sum_{i=1}^3 U_{\alpha i} \nu_{i L}$) and defining the effective mass-basis coupling vector $\lambda_{ia} \equiv \sum_{\alpha=e, \mu} y_{\alpha a} U_{\alpha i}^*$, the interaction Lagrangian is cleanly reformulated as:
\begin{equation}\label{eq:lagrangian_mass_basis}
    \mathcal{L}_{\text{mass}}^{\nu} = - \sum_{i=1}^3 \sum_{a=1,2} \left( \lambda_{ia} \bar{\nu}_i P_R N S_a + \lambda_{ia}^* \bar{N} P_L \nu_i S_a^* \right).
\end{equation}

To evaluate the loop amplitudes, the interaction states must be rotated into their physical mass bases. In the heavy fermion sector, the symmetric Majorana mass insertion $m$ combined with the Dirac mass $m_D$ inherently leads to a maximal mixing ($\theta = \pi/4$). This yields two purely Majorana mass eigenstates, $N_1$ and $N_2$, with masses $M_{1,2} = m \mp m_D$. The interaction field is thus projected as $N = \frac{1}{\sqrt{2}}(N_2 + \gamma_5 N_1)$. In the dark scalar sector, the CP-even and CP-odd components of $S_{1,2}$ mix independently. Due to their decoupled mass matrices, they rotate via generic, unequal mixing angles $\theta_R$ and $\theta_I$ into the physical mass eigenstates $H_{1,2}$ and $A_{1,2}$, respectively.

Projecting the Yukawa interactions onto these physical states, we define the effective flavor couplings between the heavy fermions $N_\alpha$ ($\alpha=1,2$) and the physical scalars $H_k, A_k$ ($k=1,2$) as:
\begin{align}
    \tilde{\lambda}_{i H_1}^* &= \lambda_{i1}^* \cos\theta_R - \lambda_{i2}^* \sin\theta_R, \nonumber \\
    \tilde{\lambda}_{i H_2}^* &= \lambda_{i1}^* \sin\theta_R + \lambda_{i2}^* \cos\theta_R, \nonumber \\
    \tilde{\lambda}_{i A_1}^* &= \lambda_{i1}^* \cos\theta_I - \lambda_{i2}^* \sin\theta_I, \nonumber \\
    \tilde{\lambda}_{i A_2}^* &= \lambda_{i1}^* \sin\theta_I + \lambda_{i2}^* \cos\theta_I.
\end{align}
The Feynman rules for the chiral $\nu_i-N_\alpha-\Phi$ vertices readily follow from these definitions, acquiring the standard left-handed projectors $P_L$ and an additional relative phase for the CP-odd scalar insertions.

Crucially, projecting the dark gauge interactions into the mass basis yields the specific vertex structures responsible for the magnetic moment. For the scalars, the misaligned mixing ($\theta_R \neq \theta_I$) dynamically induces cross-state radiative transitions. The dark photon explicitly couples CP-even and CP-odd states, yielding off-diagonal vertices such as $H_1-A_2-A'$, which carries the Feynman rule $-g_D Q^D_S \sin(\theta_R - \theta_I) (p_{\text{in}} + p_{\text{out}})^\mu$. 

In the fermion sector, the maximal mixing into Majorana eigenstates ensures that the pure vector current identically vanishes ($\overline{N_1} \gamma^\mu N_1 = 0$). Consequently, the $N-A'$ gauge interaction is strictly off-diagonal and purely axial-vector:
\begin{equation}
    \mathcal{L}_{\text{gauge}} \supset -g_D Q^D_N \overline{N_1} \gamma^\mu \gamma_5 N_2 A'_\mu,
\end{equation}
which provides the vertex factor $-i g_D Q^D_N \gamma^\mu \gamma_5$.

This architectural setup offers two profound dynamical advantages over the $\nu$SM. First, the chirality flip occurs directly on the internal heavy fermion line, fully decoupling the magnetic moment from the severe $\mathcal{O}(G_F m_\nu \cdot \Delta m^2/M_W^2)$ GIM-like suppression. Second, the dual-scalar framework furnishes independent, non-parallel flavor coupling vectors ($\vec{\lambda}_1$ and $\vec{\lambda}_2$), providing the indispensable degrees of freedom in flavor space to dynamically break the Majorana symmetry prohibition and generate a non-vanishing macroscopic transition magnetic moment. This dual-scalar topology shares profound structural similarities with radiative neutrino mass models, such as the classic Zee-Babu and scotogenic frameworks~\cite{Zee:1980ai, Krauss:2002px, Ma:2006km, Cai:2017jrq}, seamlessly connecting neutrino properties to dark sector dynamics.

\subsection{One-Loop Dynamics}

Constrained by Lorentz covariance and gauge invariance, the one-loop induced $\nu_i$-$\nu_j$-$A'$ transition matrix element is generically parameterized as:
\begin{equation}
    \mathcal{M}_{ij}^\mu = \langle \nu_j(p') | J_D^\mu | \nu_i(p) \rangle = \bar{u}_j(p') \left[ \gamma^\mu F_1(q^2) + \frac{i\sigma^{\mu\nu}q_\nu}{2m_e} F_2(q^2) + \dots \right] u_i(p).
\end{equation}
Since Majorana self-conjugacy strictly enforces a vanishing vector form factor ($F_1(q^2) \equiv 0$), the interaction is uniquely dictated by the magnetic dipole form factor $F_2(q^2)$. 

In the phenomenologically relevant regime where the momentum transfer is significantly smaller than the internal loop masses ($q^2 \ll M_{N, S}^2$), momentum-dependent corrections are kinematically suppressed. We can therefore safely evaluate the vertex in the local limit ($q^2 \to 0$), where $F_2(q^2) \simeq F_2(0)$. Thus, evaluating the physical dark transition magnetic moment $\mu_{ij}$ simply reduces to computing the one-loop amplitudes and analytically projecting out the coefficient of the $\sigma^{\mu\nu}q_\nu$ tensor structure. To facilitate this matching, we first establish the requisite Feynman rules in the mass basis.

\begin{figure}[!htbp]
    \centering

\tikzset{every picture/.style={line width=0.75pt}} 

\begin{tikzpicture}[x=0.75pt,y=0.75pt,yscale=-1,xscale=1]

\draw    (195.75,1902.52) -- (279.33,1902.62) ;
\draw [shift={(242.54,1902.57)}, rotate = 180.07] [fill={rgb, 255:red, 0; green, 0; blue, 0 }  ][line width=0.08]  [draw opacity=0] (8.93,-4.29) -- (0,0) -- (8.93,4.29) -- cycle    ;
\draw    (370.28,1903.74) -- (456.02,1904.25) ;
\draw [shift={(418.15,1904.03)}, rotate = 180.34] [fill={rgb, 255:red, 0; green, 0; blue, 0 }  ][line width=0.08]  [draw opacity=0] (8.93,-4.29) -- (0,0) -- (8.93,4.29) -- cycle    ;
\draw  [draw opacity=0] (370.28,1903.76) .. controls (370.97,1905) and (371.32,1906.28) .. (371.3,1907.6) .. controls (371.17,1917.48) and (350.2,1925.2) .. (324.46,1924.83) .. controls (298.72,1924.45) and (277.96,1916.13) .. (278.09,1906.25) .. controls (278.11,1905) and (278.46,1903.79) .. (279.1,1902.63) -- (324.7,1906.92) -- cycle ; \draw   (370.28,1903.76) .. controls (370.97,1905) and (371.32,1906.28) .. (371.3,1907.6) .. controls (371.17,1917.48) and (350.2,1925.2) .. (324.46,1924.83) .. controls (298.72,1924.45) and (277.96,1916.13) .. (278.09,1906.25) .. controls (278.11,1905) and (278.46,1903.79) .. (279.1,1902.63) ;  
\draw  [draw opacity=0][dash pattern={on 4.5pt off 4.5pt}] (279.33,1902.62) .. controls (278.64,1901.38) and (278.28,1900.09) .. (278.29,1898.78) .. controls (278.38,1888.89) and (299.31,1881.06) .. (325.05,1881.3) .. controls (350.79,1881.54) and (371.59,1889.75) .. (371.51,1899.64) .. controls (371.5,1900.89) and (371.16,1902.1) .. (370.52,1903.27) -- (324.9,1899.21) -- cycle ; \draw  [dash pattern={on 4.5pt off 4.5pt}] (279.33,1902.62) .. controls (278.64,1901.38) and (278.28,1900.09) .. (278.29,1898.78) .. controls (278.38,1888.89) and (299.31,1881.06) .. (325.05,1881.3) .. controls (350.79,1881.54) and (371.59,1889.75) .. (371.51,1899.64) .. controls (371.5,1900.89) and (371.16,1902.1) .. (370.52,1903.27) ;  
\draw  [line width=0.75] [line join = round][line cap = round] (283.85,1884.26) .. controls (283.85,1874.61) and (306.86,1872.89) .. (313.86,1872.89) ;
\draw   (306.62,1868.67) -- (313.8,1872.79) -- (306.62,1876.92) ;
\draw  [line width=0.75] [line join = round][line cap = round] (275.2,1915.26) .. controls (283.72,1921.79) and (285.92,1925.75) .. (298.04,1925.75) ;
\draw   (291.49,1922.43) -- (298.98,1925.87) -- (291.49,1929.31) ;
\draw    (326.69,1850.75) .. controls (329.38,1851.48) and (329.92,1852.83) .. (328.32,1854.82) .. controls (326.55,1856.25) and (326.4,1857.9) .. (327.89,1859.77) .. controls (329.38,1861.58) and (329.21,1863.26) .. (327.38,1864.83) .. controls (325.58,1866.38) and (325.47,1868) .. (327.04,1869.69) .. controls (328.67,1871.42) and (328.67,1873.14) .. (327.04,1874.85) .. controls (325.49,1876.58) and (325.67,1878.18) .. (327.56,1879.65) -- (328,1881.62) ;
\draw    (319.69,1869.77) -- (319.76,1855.71) ;
\draw   (322.88,1863.41) -- (319.71,1871.39) -- (316.88,1863.58) ;

\draw (152.73,1897.83) node [anchor=north west][inner sep=0.75pt]  [font=\footnotesize]  {$\nu _{i}( p)$};
\draw (467.06,1898.22) node [anchor=north west][inner sep=0.75pt]  [font=\footnotesize]  {$\nu _{j}( p')$};
\draw (314.15,1908.19) node [anchor=north west][inner sep=0.75pt]  [font=\scriptsize]  {$N_{1,2}$};
\draw (288.23,1884.4) node [anchor=north west][inner sep=0.75pt]  [font=\scriptsize]  {$S_{I,R}$};
\draw (284.01,1853) node [anchor=north west][inner sep=0.75pt]  [font=\scriptsize]  {$k$};
\draw (255.01,1919) node [anchor=north west][inner sep=0.75pt]  [font=\scriptsize]  {$p-k$};
\draw (336.23,1883.4) node [anchor=north west][inner sep=0.75pt]  [font=\scriptsize]  {$S_{R,I}$};
\draw (310.87,1827.77) node [anchor=north west][inner sep=0.75pt]  [font=\footnotesize]  {$A'( q)$};
\end{tikzpicture}
    \caption{Scalar radiation contribution diagram.}
    \label{scalarrad}
\end{figure}
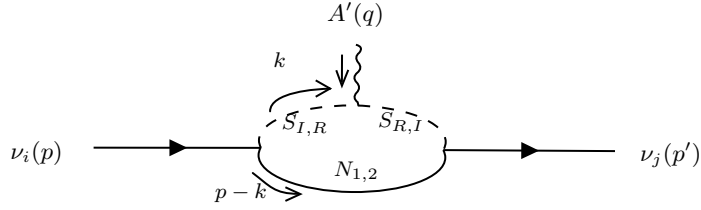

As illustrated in Fig.~\ref{ferrad} and Fig.~\ref{scalarrad}, the one-loop transition amplitudes encompass two topological classes: diagrams where the dark photon is radiated from the internal heavy fermions, and those where it is radiated from the internal dark scalars. 

\begin{figure}[!htbp]
    \centering
\tikzset{every picture/.style={line width=0.75pt}} 

\begin{tikzpicture}[x=0.75pt,y=0.75pt,yscale=-1,xscale=1]

\draw    (174.75,1641.9) -- (258.33,1642) ;
\draw [shift={(221.54,1641.95)}, rotate = 180.07] [fill={rgb, 255:red, 0; green, 0; blue, 0 }  ][line width=0.08]  [draw opacity=0] (8.93,-4.29) -- (0,0) -- (8.93,4.29) -- cycle    ;
\draw    (349.28,1643.12) -- (435.02,1643.63) ;
\draw [shift={(397.15,1643.41)}, rotate = 180.34] [fill={rgb, 255:red, 0; green, 0; blue, 0 }  ][line width=0.08]  [draw opacity=0] (8.93,-4.29) -- (0,0) -- (8.93,4.29) -- cycle    ;
\draw  [draw opacity=0] (349.28,1643.14) .. controls (349.97,1644.38) and (350.32,1645.66) .. (350.3,1646.98) .. controls (350.17,1656.87) and (329.2,1664.58) .. (303.46,1664.21) .. controls (277.72,1663.83) and (256.96,1655.52) .. (257.09,1645.63) .. controls (257.11,1644.38) and (257.46,1643.17) .. (258.1,1642.01) -- (303.7,1646.3) -- cycle ; \draw   (349.28,1643.14) .. controls (349.97,1644.38) and (350.32,1645.66) .. (350.3,1646.98) .. controls (350.17,1656.87) and (329.2,1664.58) .. (303.46,1664.21) .. controls (277.72,1663.83) and (256.96,1655.52) .. (257.09,1645.63) .. controls (257.11,1644.38) and (257.46,1643.17) .. (258.1,1642.01) ;  
\draw  [draw opacity=0][dash pattern={on 4.5pt off 4.5pt}] (258.2,1641.78) .. controls (257.52,1640.54) and (257.16,1639.26) .. (257.17,1637.95) .. controls (257.25,1628.05) and (278.18,1620.23) .. (303.93,1620.46) .. controls (329.67,1620.7) and (350.47,1628.91) .. (350.39,1638.81) .. controls (350.38,1640.05) and (350.04,1641.26) .. (349.4,1642.43) -- (303.78,1638.38) -- cycle ; \draw  [dash pattern={on 4.5pt off 4.5pt}] (258.2,1641.78) .. controls (257.52,1640.54) and (257.16,1639.26) .. (257.17,1637.95) .. controls (257.25,1628.05) and (278.18,1620.23) .. (303.93,1620.46) .. controls (329.67,1620.7) and (350.47,1628.91) .. (350.39,1638.81) .. controls (350.38,1640.05) and (350.04,1641.26) .. (349.4,1642.43) ;  
\draw    (314.85,1679.7) -- (315.11,1695.43) ;
\draw   (311.32,1687.89) -- (314.81,1678.88) -- (318.28,1687.53) ;
\draw  [line width=0.75] [line join = round][line cap = round] (262.85,1623.65) .. controls (262.85,1613.99) and (285.86,1612.27) .. (292.86,1612.27) ;
\draw   (285.62,1608.05) -- (292.8,1612.18) -- (285.62,1616.3) ;
\draw  [line width=0.75] [line join = round][line cap = round] (254.2,1654.64) .. controls (262.72,1661.17) and (264.92,1665.14) .. (277.04,1665.14) ;
\draw   (270.49,1661.81) -- (277.98,1665.25) -- (270.49,1668.69) ;
\draw    (303.69,1664.13) .. controls (306.38,1664.86) and (306.92,1666.21) .. (305.32,1668.2) .. controls (303.55,1669.63) and (303.4,1671.29) .. (304.89,1673.16) .. controls (306.38,1674.97) and (306.21,1676.65) .. (304.38,1678.21) .. controls (302.58,1679.77) and (302.47,1681.39) .. (304.04,1683.07) .. controls (305.67,1684.8) and (305.67,1686.52) .. (304.04,1688.23) .. controls (302.49,1689.96) and (302.67,1691.56) .. (304.56,1693.03) -- (305,1695) ;

\draw (131.73,1637.22) node [anchor=north west][inner sep=0.75pt]  [font=\footnotesize]  {$\nu _{i}( p)$};
\draw (446.06,1637.6) node [anchor=north west][inner sep=0.75pt]  [font=\footnotesize]  {$\nu _{j}( p')$};
\draw (265.15,1641.57) node [anchor=north west][inner sep=0.75pt]  [font=\scriptsize]  {$N_{1,2}$};
\draw (307.23,1602.78) node [anchor=north west][inner sep=0.75pt]  [font=\scriptsize]  {$S_{I,R}$};
\draw (286.87,1695.77) node [anchor=north west][inner sep=0.75pt]  [font=\footnotesize]  {$A'( q)$};
\draw (263.01,1592.38) node [anchor=north west][inner sep=0.75pt]  [font=\scriptsize]  {$k$};
\draw (234.01,1657.38) node [anchor=north west][inner sep=0.75pt]  [font=\scriptsize]  {$p-k$};
\draw (318.15,1639.57) node [anchor=north west][inner sep=0.75pt]  [font=\scriptsize]  {$N_{2,1}$};

\end{tikzpicture}
    \caption{Fermion radiation contribution diagram.}
    \label{ferrad}
\end{figure}
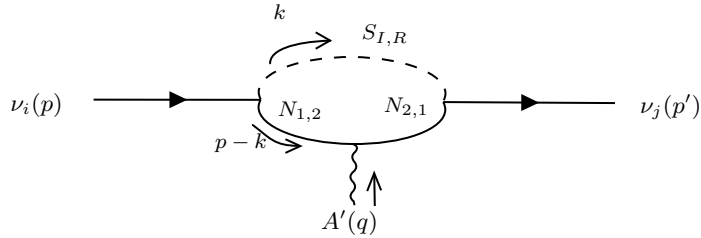

We first evaluate the fermion-radiated contributions. For a given internal physical scalar $\Phi \in \{H_k, A_k\}$ and a specific fermion transition $N_1 \to N_2$, the loop amplitude reads:
\begin{equation}
    i\mathcal{M}^\mu_{\Phi, 12} = C_{\Phi ij} \int \frac{d^4k}{(2\pi)^4} \frac{\overline{u}_{j}(p') P_{L} \left[ (\not{p}' - \not{k} + M_2) \gamma^\mu \gamma_5 (\not{p} - \not{k} + M_1) \right] P_{L} u_{i}(p)}{[(p'-k)^2 - M_2^2][(p-k)^2 - M_1^2][k^2 - M_\Phi^2]},
\end{equation}
where $C_{\Phi ij}$ encapsulates the coupling factors, which carry a relative sign difference depending on the CP parity of $\Phi$ due to the scalar field expansion. 

To isolate the magnetic dipole structure, we introduce standard Feynman parameters ($x, y$) and shift the loop momentum $l^\mu = k^\mu - x p'^\mu - y p^\mu$. Working in the limit of minimal momentum transfer ($q^2 \to 0$), the chiral projectors $P_L (\dots) P_L$ strictly filter out terms lacking a heavy mass insertion, successfully satisfying the chirality flip requirement. Extracting the coefficient of the $i\sigma^{\mu\nu}q_\nu$ tensor structure and completing the momentum integration yields the single-direction contribution:
\begin{equation}
    \mu_{L, 12}^{(\Phi)} = \frac{C_{\Phi ij}}{16 \pi^2} \int_0^1 dx \int_0^{1-x} dy \frac{M_2(1+x-y) - M_1(1-x+y)}{\Delta_{\Phi, 12}(x, y)},
\end{equation}
where the effective denominator is $\Delta_{\Phi, 12}(x, y) = x M_2^2 + y M_1^2 + (1-x-y) M_\Phi^2$.

The total physical contribution from this topology must include the reverse transition, $N_2 \to N_1$. By gauge symmetry, its amplitude $\mu_{L, 21}^{(\Phi)}$ is obtained simply by exchanging the internal mass parameters $M_1 \leftrightarrow M_2$:
\begin{equation}
    \mu_{L, 21}^{(\Phi)} = \frac{C_{\Phi ij}}{16 \pi^2} \int_0^1 dx \int_0^{1-x} dy \frac{M_1(1+x-y) - M_2(1-x+y)}{\Delta_{\Phi, 21}(x, y)},
\end{equation}
with $\Delta_{\Phi, 21}(x, y) = x M_1^2 + y M_2^2 + (1-x-y) M_\Phi^2$. Performing the variable substitution $x \leftrightarrow y$, the symmetric integration domain and denominator ($\Delta_{\Phi, 21}(y, x) \equiv \Delta_{\Phi, 12}(x, y)$) remain invariant. The numerator, however, is strictly antisymmetric under this exchange:
\begin{equation}
    M_1(1+y-x) - M_2(1-y+x) = - \left[ M_2(1+x-y) - M_1(1-x+y) \right].
\end{equation}
This establishes the exact analytical relation $\mu_{L, 21}^{(\Phi)} = - \mu_{L, 12}^{(\Phi)}$.

This exact cancellation is not a numerical coincidence, but a fundamental dynamical consequence of Majorana self-conjugacy coupled with the purely axial-vector structure ($\gamma^\mu \gamma_5$) of the dark gauge interaction in the physical basis. Summing over all internal states, the net contribution from fermion-radiated diagrams vanishes identically ($\mu_{L}^{\text{Fermion-Rad}} \equiv 0$). Therefore, the generation of a macroscopic dark transition magnetic moment relies entirely on the scalar-radiated topologies.

As demonstrated above, the physical origin of the dark transition magnetic moment is entirely dominated by scalar-radiated topologies (Fig.~\ref{scalarrad}), where the dark photon $A'_\mu$ couples directly to the internal scalar propagators.

The foundational gauge coupling arises from the scalar kinetic term: $$\mathcal{L}_{S-A'} = - g_D Q^D_S A'^\mu \sum_{a=1,2} S_{aR} \overleftrightarrow{\partial_\mu} S_{aI}.$$ Rotating the interaction states $(S_{aR}, S_{aI})$ into the physical mass bases $(H_k, A_m)$ governed by their independent mixing angles ($\theta_R$ and $\theta_I$), the physical gauge interactions analytically condense to:
\begin{align}
\mathcal{L}_{S-A'} = - g_D Q^D_S A'^\mu \Big[ & \cos(\theta_R - \theta_I) \left( H_1 \overleftrightarrow{\partial_\mu} A_1 + H_2 \overleftrightarrow{\partial_\mu} A_2 \right) \nonumber \\
+ & \sin(\theta_I - \theta_R) \left( H_1 \overleftrightarrow{\partial_\mu} A_2 - H_2 \overleftrightarrow{\partial_\mu} A_1 \right) \Big].
\end{align}
The antisymmetric derivative fundamentally prohibits same-parity couplings (e.g., $H_i \overleftrightarrow{\partial_\mu} H_j$). More importantly, the misaligned mixing ($\theta_R \neq \theta_I$) dynamically induces off-diagonal cross-state transitions between intergenerational scalars (with coefficients proportional to $\pm\sin(\theta_I - \theta_R)$).

For a given internal heavy fermion $N_\alpha$ ($\alpha=1,2$) and a specific scalar transition path $\Phi_1 \to \Phi_2 A'$, the loop amplitude reads:
\begin{equation}
i\mathcal{M}^\mu_{\Phi_1 \Phi_2, \alpha} = \int \frac{d^4k}{(2\pi)^4} \overline{u}_j(p') \left(\frac{\tilde{\lambda}_{j \Phi_2}^*}{2} P_R\right) \frac{i(\not{k} + M_\alpha)}{k^2 - M_\alpha^2} \left(\frac{\tilde{\lambda}_{i \Phi_1}^*}{2} P_L\right) u_i(p) \frac{i V^\mu}{D_{\Phi_2} D_{\Phi_1}},
\end{equation}
where $V^\mu = -i g_D Q^D_S C_{\Phi_1 \Phi_2} (p + p' - 2k)^\mu$ is the gauge vertex extracted from $\mathcal{L}_{S-A'}$ with the corresponding trigonometric coefficient $C_{\Phi_1 \Phi_2}$. 

Evaluating the numerator trace, the chiral projectors strictly enforce $P_R (\not{k} + M_\alpha) P_L = M_\alpha P_R$, effectively filtering out the momentum terms and isolating the internal mass insertion required for the chirality flip. Applying the Gordon identity and projecting out the $i\sigma^{\mu\nu}q_\nu$ tensor structure, the magnetic moment contribution from a single path, parameterized in units of the Bohr magneton ($\mu_B = e/2m_e$), is compactly expressed as:
\begin{equation}
\mu_{ij}^{(\Phi_1 \Phi_2, \alpha)} = \frac{g_D Q^D_S}{32 \pi^2 e} \left( m_e M_\alpha \right) C_{\Phi_1 \Phi_2} \left( \tilde{\lambda}_{j \Phi_2}^* \tilde{\lambda}_{i \Phi_1}^* \right) \mathcal{J}(M_\alpha, M_{\Phi_1}, M_{\Phi_2}) \mu_B,
\end{equation}
where the dimensionless loop integral function is:
\begin{equation}
\mathcal{J}(M_\alpha, M_{\Phi_1}, M_{\Phi_2}) = \int_0^1 dx \int_0^{1-x} dy \frac{x+y}{x M_{\Phi_2}^2 + y M_{\Phi_1}^2 + (1-x-y) M_\alpha^2}.
\end{equation}

The total physical TMM requires a coherent summation over all allowed scalar transition paths and internal fermions $N_\alpha$, followed by a flavor space antisymmetrization ($\mathcal{A}_{ij} = \mu_{ij} - \mu_{ji}$). Diagonal flavor structures inherently cancel under this operation. The resulting analytical expression for the macroscopic dark transition magnetic moment simplifies beautifully to:
\begin{equation}\label{eq:final_magnetic_moment}
\mu_{ij} = \frac{g_D Q^D_S}{64 \pi^2} \sin(2(\theta_I - \theta_R)) \cdot \text{Im}(\lambda_{j2}^* \lambda_{i1}^* - \lambda_{i2}^* \lambda_{j1}^*) \sum_{\alpha=1,2} M_\alpha \Delta \mathcal{J}_\alpha,
\end{equation}
with the generalized mass splitting factor defined as:
\begin{equation}
\Delta \mathcal{J}_\alpha = \mathcal{J}(M_\alpha, M_{H_1}, M_{A_1}) + \mathcal{J}(M_\alpha, M_{H_2}, M_{A_2}) - \mathcal{J}(M_\alpha, M_{H_1}, M_{A_2}) - \mathcal{J}(M_\alpha, M_{H_2}, M_{A_1}).
\end{equation}

Equation~\eqref{eq:final_magnetic_moment} dictates that evading the severe prohibitions of single-generation models requires the simultaneous fulfillment of three distinct dynamical conditions. Specifically, the survival of the macroscopic transition magnetic moment relies on a flavor-space misalignment---characterized by non-parallel Yukawa coupling vectors yielding a non-vanishing antisymmetric tensor $(\lambda_{j2}^* \lambda_{i1}^* - \lambda_{i2}^* \lambda_{j1}^*) \neq 0$---coupled with an internal mixing misalignment between the CP-even and CP-odd scalar components ($\sin(2(\theta_I - \theta_R)) \neq 0$) that fundamentally induces the off-diagonal gauge currents. Furthermore, these symmetry-breaking structures must be accompanied by a non-degenerate physical scalar mass spectrum ($\Delta \mathcal{J}_\alpha \neq 0$), without which a GIM-like mechanism would enforce an exact cancellation of the integral structure.

Before proceeding to the global phenomenological analysis, a brief theoretical remark regarding the structural correlation between the dark transition magnetic moment and the radiative neutrino mass is warranted. In generic radiative frameworks, removing the external gauge boson line from the magnetic moment loop invariably induces a radiative correction to the neutrino mass, denoted as $\delta m_\nu$. As famously pointed out by Voloshin~\cite{Voloshin:1987qy} and comprehensively analyzed by Bell et al.~\cite{Bell:2005kz}, generating a macroscopic magnetic moment ($\mu_\nu \sim 10^{-11} \mu_B$) typically implies an unacceptably large $\delta m_\nu \sim (\Lambda^2 / 2 m_e) \mu_\nu$. Satisfying the stringent sub-eV cosmological mass bounds therefore seemingly demands extreme parameter fine-tuning, unless an approximate symmetry (such as an $SU(2)_V$ spin-flavor symmetry) is explicitly invoked.

However, it is crucial to recognize the fundamental quantum field-theoretic distinction between these two quantities. The transition magnetic moment originates from a dimension-five operator that yields a strictly finite, calculable prediction at the one-loop level. In contrast, the radiative mass correction constitutes a lower-dimensional operator that is generically ultraviolet-divergent. Consequently, $\delta m_\nu$ necessitates rigorous renormalization and is ultimately absorbed into the bare mass parameters of the fundamental theory. Because their underlying generation mechanisms and ultraviolet sensitivities are intrinsically distinct, the mass and the magnetic moment do not obey an inextirpable, rigid proportionality. 

While dynamically resolving this naturalness tension typically requires embedding the current framework into a more elaborate symmetry-protected ultraviolet completion, the primary objective of this work is to isolate and demonstrate the phenomenological dominance of indirect high-energy constraints (cLFV and accelerator limits) over direct scattering bounds. We therefore acknowledge the requisite fine-tuning in the mass sector within this simplified dual-scalar topology, but deliberately defer the construction of a fully natural, symmetry-protected model to future investigations.

\section{Global Constraints}
\label{sec:global_constraints}
\subsection{The Visible Sector: $\mu \to e \gamma$}
\label{subsec:visible_constraints}
To robustly validate the misaligned double-scalar framework, its theoretical predictions must be confronted with a comprehensive suite of experimental constraints. The survival of a macroscopic dark magnetic moment relies on a delicate interplay of parameters that are simultaneously subjected to stringent, multi-frontier limits. In this section, we systematically evaluate the boundaries imposed by visible-sector observables and dark-sector accelerator probes, revealing a profound physical tension that dynamically shapes the viable parameter space.

The heavy charged fermion $E$ and the dark scalars $S_a$ inevitably induce charged lepton flavor violation in the visible sector~\cite{Lindner:2016lxq, Calibbi:2017uvl}. Because the identical Yukawa interactions responsible for the dark magnetic moment also mediate these visible transitions, the radiative decay $\mu \to e \gamma$ currently serves as the most stringent experimental probe of this framework. Furthermore, this visible-sector boundary will be subjected to even more rigorous tests by upcoming experiments like Mu3e, Mu2e, and COMET, which are poised to probe significantly deeper regions of the parameter space~\cite{Mu3e:2020gyw, Mu2e:2014fns, COMET:2018auw}.

The relevant flavor-violating interactions originate from the Yukawa Lagrangian:
\begin{equation}
    \mathcal{L}_{\text{int}} \supset - \sum_{a=1,2} \left( y_{ea} \bar{e}_L E_R S_a + y_{\mu a} \bar{\mu}_L E_R S_a + \text{h.c.} \right).
\end{equation}
Projecting the scalars into their physical mass eigenstates $(H_k, A_m)$, the effective couplings $\tilde{y}_{\alpha \Phi}$ ($\Phi \in \{H_k, A_m\}$) are dynamically governed by the respective scalar mixing angles $\theta_R$ and $\theta_I$.

\begin{figure}[!htbp]
    \centering

\tikzset{every picture/.style={line width=0.75pt}} 

\begin{tikzpicture}[x=0.75pt,y=0.75pt,yscale=-1,xscale=1]

\draw    (179.29,767.66) -- (271.85,767.77) ;
\draw [shift={(230.57,767.72)}, rotate = 180.07] [fill={rgb, 255:red, 0; green, 0; blue, 0 }  ][line width=0.08]  [draw opacity=0] (8.93,-4.29) -- (0,0) -- (8.93,4.29) -- cycle    ;
\draw    (372.81,769.06) -- (467.75,769.63) ;
\draw [shift={(425.28,769.38)}, rotate = 180.35] [fill={rgb, 255:red, 0; green, 0; blue, 0 }  ][line width=0.08]  [draw opacity=0] (8.93,-4.29) -- (0,0) -- (8.93,4.29) -- cycle    ;
\draw    (321.52,794.82) .. controls (324.4,794.89) and (325.23,796.09) .. (324.02,798.44) .. controls (322.37,799.99) and (322.34,801.66) .. (323.91,803.44) .. controls (325.44,805.19) and (325.28,806.86) .. (323.45,808.43) .. controls (321.62,809.89) and (321.43,811.57) .. (322.9,813.46) .. controls (324.39,815.19) and (324.23,816.77) .. (322.4,818.2) .. controls (320.57,819.91) and (320.41,821.67) .. (321.93,823.48) .. controls (323.48,825.26) and (323.37,826.94) .. (321.62,828.51) .. controls (319.9,830.15) and (319.87,831.82) .. (321.52,833.51) .. controls (323.22,835.1) and (323.28,836.72) .. (321.7,838.39) .. controls (320.18,840.15) and (320.37,841.81) .. (322.27,843.36) -- (323.08,847.05) ;
\draw  [draw opacity=0] (372.87,769.17) .. controls (373.59,770.55) and (373.96,771.98) .. (373.94,773.43) .. controls (373.8,784.65) and (350.57,793.41) .. (322.07,792.99) .. controls (293.57,792.57) and (270.58,783.12) .. (270.72,771.9) .. controls (270.74,770.52) and (271.11,769.18) .. (271.79,767.89) -- (322.33,772.66) -- cycle ; \draw   (372.87,769.17) .. controls (373.59,770.55) and (373.96,771.98) .. (373.94,773.43) .. controls (373.8,784.65) and (350.57,793.41) .. (322.07,792.99) .. controls (293.57,792.57) and (270.58,783.12) .. (270.72,771.9) .. controls (270.74,770.52) and (271.11,769.18) .. (271.79,767.89) ;  
\draw  [draw opacity=0][dash pattern={on 4.5pt off 4.5pt}] (271.78,767.65) .. controls (271.06,766.27) and (270.68,764.85) .. (270.69,763.39) .. controls (270.78,752.16) and (293.96,743.28) .. (322.47,743.55) .. controls (350.97,743.82) and (374.01,753.15) .. (373.92,764.39) .. controls (373.91,765.76) and (373.55,767.11) .. (372.87,768.4) -- (322.31,763.89) -- cycle ; \draw  [dash pattern={on 4.5pt off 4.5pt}] (271.78,767.65) .. controls (271.06,766.27) and (270.68,764.85) .. (270.69,763.39) .. controls (270.78,752.16) and (293.96,743.28) .. (322.47,743.55) .. controls (350.97,743.82) and (374.01,753.15) .. (373.92,764.39) .. controls (373.91,765.76) and (373.55,767.11) .. (372.87,768.4) ;  
\draw    (345.84,791.37) -- (334.46,793.17) ;
\draw [shift={(346.57,791.25)}, rotate = 171.01] [fill={rgb, 255:red, 0; green, 0; blue, 0 }  ][line width=0.08]  [draw opacity=0] (8.93,-4.29) -- (0,0) -- (8.93,4.29) -- cycle    ;
\draw  [line width=0.75] [line join = round][line cap = round] (276.54,746.2) .. controls (289.23,733.13) and (301.2,733.91) .. (319.81,733.91) ;
\draw   (279.42,737.6) -- (274.88,748.92) -- (285.65,743.7) ;
\draw    (334.02,814.49) -- (334.4,835.92) ;
\draw   (329.03,825.66) -- (333.97,813.38) -- (338.87,825.17) ;

\draw (136.41,758.54) node [anchor=north west][inner sep=0.75pt]  [font=\small]  {$\mu ^{-}( p)$};
\draw (471.17,760.51) node [anchor=north west][inner sep=0.75pt]  [font=\small]  {$e^{-}( p')$};
\draw (358.45,729.28) node [anchor=north west][inner sep=0.75pt]  [font=\small]  {$\Phi $};
\draw (271.7,786.54) node [anchor=north west][inner sep=0.75pt]  [font=\small]  {$E^{-}$};
\draw (323.59,837.64) node [anchor=north west][inner sep=0.75pt]  [font=\small]  {$\gamma ( q)$};
\draw (280.4,714.19) node [anchor=north west][inner sep=0.75pt]  [font=\small]  {$k$};

\end{tikzpicture}
    \caption{One-loop Feynman diagram inducing $\mu^- \to e^- \gamma$.}
    \label{mu_to_egamma_loop}
\end{figure}
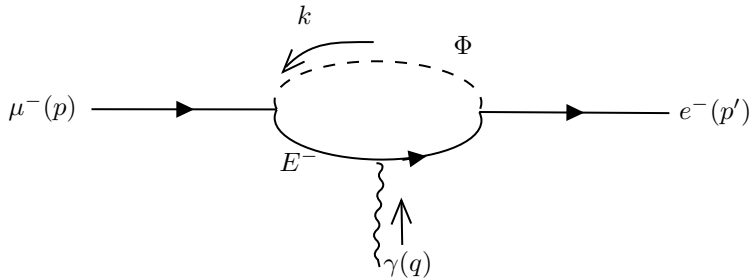

At the one-loop level, the $\mu \to e \gamma$ transition is mediated by the internal charged fermion $E$ and the neutral scalars $\Phi$, as illustrated in Fig.~\ref{mu_to_egamma_loop}. Matching the loop amplitudes to the effective electromagnetic dipole operator $\mathcal{L}_{\text{dipole}} = - \frac{e m_\mu}{2} \bar{e} \sigma^{\rho\nu} P_R \mu F_{\rho\nu} + \text{h.c.}$, we analytically extract the total right-handed Wilson coefficient:
\begin{equation}
    A_R = \frac{1}{32 \pi^2 M_E^2} \sum_{\Phi \in \{H_k, A_m\}} \tilde{y}_{e \Phi} \tilde{y}_{\mu \Phi}^* \tilde{F}_2\left( x_\Phi \right),
\end{equation}
where $x_\Phi \equiv M_\Phi^2 / M_E^2$, and the standard loop function is defined as~\cite{Lavoura_2003}:
\begin{equation}
    \tilde{F}_2(x) = \frac{2 + 3x - 6x^2 + x^3 + 6x \ln x}{6(1-x)^4}.
\end{equation}

An elegantly exact theoretical simplification emerges in the heavy fermion decoupling limit ($M_E \gg M_\Phi$). In this kinematic regime, the loop function smoothly converges to $\tilde{F}_2(0) = 1/3$. Crucially, enforcing the unitarity of the orthogonal rotation matrices ($\cos^2\theta_{R,I} + \sin^2\theta_{R,I} = 1$), the summation over the physical mass eigenstates collapses analytically back to the original interaction-basis couplings:
\begin{equation}
    \sum_{k=1,2} \tilde{y}_{e H_k} \tilde{y}_{\mu H_k}^* = \sum_{a=1,2} y_{ea} y_{\mu a}^*, \quad \sum_{m=1,2} \tilde{y}_{e A_m} \tilde{y}_{\mu A_m}^* = \sum_{a=1,2} y_{ea} y_{\mu a}^*.
\end{equation}
Since the vertices of the CP-odd scalars inherently carry an additional factor of $i$ from the complex expansion, their amplitudes constructively interfere with those of the CP-even scalars. Consequently, regardless of the complex mass splittings and misaligned mixings in the dark sector, the total dipole amplitude in the heavy scale limit robustly scales with the total effective Yukawa couplings:
\begin{equation}\label{eq:AR_limit}
    A_R \simeq \frac{1}{48 \pi^2 M_E^2} \sum_{a=1,2} y_{ea} y_{\mu a}^*.
\end{equation}
This demonstrates that the $\mu \to e \gamma$ observable effectively shields the dark scalar mixing parameters, isolating the fundamental flavor-violating strength of the theory.

Neglecting the electron mass, the two-body partial width for the radiative decay is $\Gamma(\mu \to e \gamma) \simeq \frac{\alpha_{\text{em}} m_\mu^5}{8} |A_R|^2$. Normalizing this against the total tree-level muon decay width $\Gamma_{\text{total}} \simeq \frac{G_F^2 m_\mu^5}{192 \pi^3}$, the closed-form branching ratio predicted by our model reduces to:
\begin{equation}
    \text{BR}(\mu \to e \gamma) = \frac{24 \pi^3 \alpha_{\text{em}}}{G_F^2} |A_R|^2 \simeq \frac{\alpha_{\text{em}}}{96 \pi G_F^2 M_E^4} \left| \sum_{a=1,2} y_{ea} y_{\mu a}^* \right|^2.
\end{equation}

This prediction must accommodate the extraordinarily stringent upper bound recently established by the MEG II collaboration~\cite{megiicollaboration2024searchmutoegammadataset}:
\begin{equation}
    \text{BR}(\mu \to e \gamma) < 3.1 \times 10^{-13} \quad (90\% \text{ C.L.}).
\end{equation}
Setting our benchmark heavy fermion scale to $M_E = 2~\text{TeV}$, the MEG II exclusion line imposes a severe numerical boundary on the original flavor-violating parameter space:
\begin{equation}
    \sqrt{\left| \sum_{a=1,2} y_{ea} y_{\mu a}^* \right|} \lesssim 0.086 \times \left( \frac{M_E}{2~\text{TeV}} \right).
\end{equation}
This robust $\mathcal{O}(10^{-2})$ constraint dictates that the identical Yukawa couplings required to generate a macroscopic dark transition magnetic moment are strictly confined to the deep perturbative regime. This stringent visible-sector boundary fundamentally shapes the viable parameter space and dynamically induces the profound physical tension highlighted in our subsequent global phenomenological analysis.

At the LHC, the heavy vector-like lepton $E^\pm$ is predominantly pair-produced via the electroweak Drell-Yan process and subsequently decays into a Standard Model charged lepton and a dark scalar ($E^\pm \to \ell^\pm S_a$). The resulting opposite-sign dilepton plus missing transverse energy ($\ell^+\ell^- + E_T^{\text{miss}}$) signature is severely constrained by recent dedicated searches from the CMS~\cite{PhysRevD.105.112007} and ATLAS~\cite{ATLAS:2024mrr} collaborations using Run 2 data, which exclude such vector-like leptons coupled to first- and second-generation leptons up to approximately $1.27~\text{TeV}$.

Recent ATLAS and CMS searches for vector-like leptons in the $\ell^+\ell^- + E_T^{\text{miss}}$ channel exclude masses up to approximately $1.27~\text{TeV}$~\cite{PhysRevD.105.112007, ATLAS:2024mrr}. To safely accommodate these direct collider limits while simultaneously addressing the severe cLFV constraints derived above, we decouple the heavy charged fermion at a benchmark scale of $M_E = 2~\text{TeV}$. This choice seamlessly evades current LHC exclusions and provides sufficient $1/M_E^4$ kinematic suppression to the $\mu \to e \gamma$ amplitude, ensuring that the requisite flavor-violating Yukawa couplings remain strictly perturbative for the subsequent phenomenological analysis.

The identical one-loop topologies mediating $\mu \to e \gamma$ inherently generate a flavor-conserving anomalous magnetic moment for the muon ($\Delta a_\mu$). Substituting the diagonal couplings into the asymptotic dipole limit yields:
\begin{equation}
    \Delta a_\mu \simeq \frac{m_\mu^2}{24 \pi^2 M_E^2} \sum_{a=1,2} |y_{\mu a}|^2.
\end{equation}
While recent Fermilab measurements report an $\mathcal{O}(10^{-9})$ deviation from the SM~\cite{Aoyama:2020ynm, Muong-2:2023cdq}, accommodating this anomaly is structurally incompatible with our framework. Barring highly fine-tuned flavor hierarchies ($y_{ea} \ll y_{\mu a}$), the exceptionally stringent MEG II bound on the off-diagonal couplings ($\sqrt{|y_{ea} y_{\mu a}^*|} \lesssim \mathcal{O}(10^{-2})$) robustly restricts the diagonal components. Under natural parameter configurations ($y_{ea} \sim y_{\mu a}$), the dynamically induced $\Delta a_\mu$ is severely suppressed to $\mathcal{O}(10^{-13})$. Consequently, the $(g-2)_\mu$ correction is safely decoupled and completely eclipsed by experimental uncertainties, cementing cLFV as the overwhelmingly dominant constraint on the visible sector.

\subsection{The Dark Sector: Borexino and NA64}
\label{subsec:dark_constraints}

To rigorously evaluate the viability of the dark transition magnetic moment within our framework, it is imperative to cross-examine the macroscopic observable $\mu_{\text{eff}} = \epsilon \cdot \mu_{ij}$ against both astrophysical direct detection limits and accelerator-based indirect bounds. 

\begin{figure}[!htbp]
    \centering

\tikzset{every picture/.style={line width=0.75pt}} 

\begin{tikzpicture}[x=0.75pt,y=0.75pt,yscale=-1,xscale=1]

\draw    (474.19,1043.01) -- (376.6,1042.64) ;
\draw [shift={(431.89,1042.85)}, rotate = 180.22] [fill={rgb, 255:red, 0; green, 0; blue, 0 }  ][line width=0.08]  [draw opacity=0] (8.93,-4.29) -- (0,0) -- (8.93,4.29) -- cycle    ;
\draw    (376.6,1042.64) -- (286,1043.59) ;
\draw [shift={(337.8,1043.05)}, rotate = 179.39] [fill={rgb, 255:red, 0; green, 0; blue, 0 }  ][line width=0.08]  [draw opacity=0] (8.93,-4.29) -- (0,0) -- (8.93,4.29) -- cycle    ;
\draw   (379.48,972.93) .. controls (382.92,975.01) and (386.2,977.01) .. (386.23,979.37) .. controls (386.27,981.74) and (383.06,983.82) .. (379.69,986) .. controls (376.31,988.18) and (373.1,990.27) .. (373.14,992.64) .. controls (373.18,995) and (376.46,996.99) .. (379.9,999.08) .. controls (383.34,1001.16) and (386.62,1003.16) .. (386.66,1005.52) .. controls (386.7,1007.89) and (383.49,1009.97) .. (380.11,1012.15) .. controls (376.74,1014.33) and (373.53,1016.42) .. (373.57,1018.79) .. controls (373.6,1021.15) and (376.88,1023.14) .. (380.32,1025.23) .. controls (383.77,1027.31) and (387.04,1029.31) .. (387.08,1031.67) .. controls (387.12,1034.04) and (383.91,1036.12) .. (380.54,1038.3) .. controls (378.91,1039.35) and (377.33,1040.38) .. (376.12,1041.43) ;
\draw    (474.19,973.27) -- (376.6,972.9) ;
\draw [shift={(431.89,973.11)}, rotate = 180.22] [fill={rgb, 255:red, 0; green, 0; blue, 0 }  ][line width=0.08]  [draw opacity=0] (8.93,-4.29) -- (0,0) -- (8.93,4.29) -- cycle    ;
\draw    (376.6,972.9) -- (285.46,973.71) ;
\draw [shift={(337.53,973.24)}, rotate = 179.49] [fill={rgb, 255:red, 0; green, 0; blue, 0 }  ][line width=0.08]  [draw opacity=0] (8.93,-4.29) -- (0,0) -- (8.93,4.29) -- cycle    ;

\draw (257.98,966.02) node [anchor=north west][inner sep=0.75pt]  [font=\LARGE]  {$^{\nu _{e}}$};
\draw (483.03,966.02) node [anchor=north west][inner sep=0.75pt]  [font=\LARGE]  {$^{\nu _{\mu }}$};
\draw (260.07,1033.9) node [anchor=north west][inner sep=0.75pt]  [font=\LARGE]  {$^{e^{-}}$};
\draw (485.3,1035.34) node [anchor=north west][inner sep=0.75pt]  [font=\LARGE]  {$^{e^{-}}$};
\draw (394.28,1001.24) node [anchor=north west][inner sep=0.75pt]  [font=\large]  {$^{A'}$};

\end{tikzpicture}
    \caption{Diagram of dark photon-mediated neutrino-electron elastic scattering.}
    \label{fig:scattering_diagram}
\end{figure}
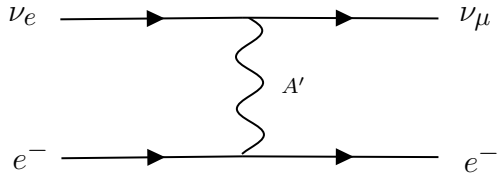

We begin by assessing the direct constraints from solar neutrino scattering. The Borexino experiment provides the leading bounds on the effective magnetic moment via $\nu e^- \to \nu e^-$ elastic scattering. As illustrated in Fig.~\ref{fig:scattering_diagram}, this macroscopic process is dynamically mediated in our model by the $t$-channel exchange of the massive dark photon $A'$. 

Assuming a standard massless photon mediator, the Borexino collaboration originally reported a baseline limit of $\mu_{\text{Borexino}}^{\text{limit}} = 2.8 \times 10^{-11} \mu_B$ (90\% C.L.) by observing no excess events in the low recoil energy region~\cite{Agostini_2017}. To rigorously map this experimental observation onto a model containing a massive $A'$, data recasting is required based on the principle of event rate conservation within the detector's specific recoil energy window $[T_{\text{min}}, T_{\text{max}}]$:
\begin{equation}
    (\mu_{\text{eff}}^{\text{limit}})^2 \int_{T_{\text{min}}}^{T_{\text{max}}} \left( \frac{1}{T_e} - \frac{1}{E_\nu} \right) \mathcal{F}^2(T_e, M_{A'}) \, dT_e = (\mu_{\text{Borexino}}^{\text{limit}})^2 \int_{T_{\text{min}}}^{T_{\text{max}}} \left( \frac{1}{T_e} - \frac{1}{E_\nu} \right) dT_e,
\end{equation}
where $\mathcal{F}(T_e, M_{A'})$ is the dynamical shape factor accounting for the modification of the mediator propagator. For a target electron at rest, the squared four-momentum transfer in the $t$-channel is given by $q^2 = -2m_e T_e$. The transition from a massless photon to a massive dark photon alters the scattering amplitude by the ratio of their propagators, uniquely determining the shape factor as:
\begin{equation}
    \mathcal{F}(T_e, M_{A'}) = \frac{q^2}{q^2 - M_{A'}^2} = \frac{2m_e T_e}{2m_e T_e + M_{A'}^2}.
\end{equation}

\begin{figure}[!htbp]
    \centering
    \includegraphics[width=0.85\textwidth]{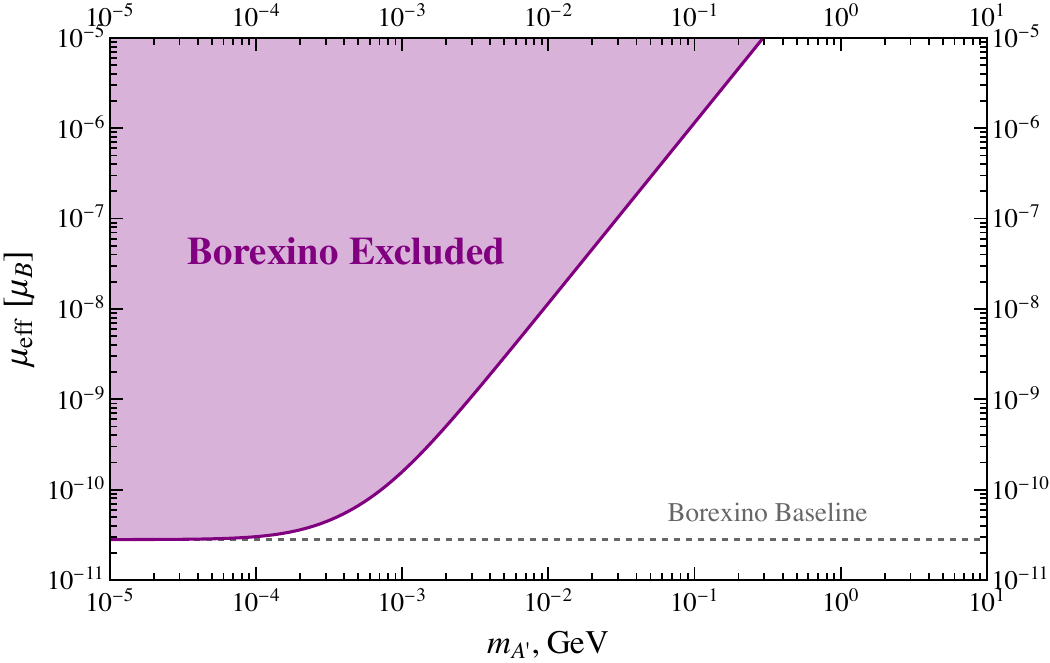} 
    \caption{Independent exclusion limits from the Borexino experiment~\cite{Agostini_2017} on the effective neutrino dark magnetic moment $\mu_{\text{eff}}$ (90\% C.L.). The solid purple line illustrates the recast limits incorporating dark photon propagator mass effects scaling with $M_{A'}$; the dashed gray line represents the baseline limit under the standard massless photon assumption.}
    \label{fig:borexino_limit}
\end{figure}

Substituting Borexino's recoil energy threshold $T_{\text{min}} = 50 \text{ keV}$ and the monochromatic energy of $^7\text{Be}$ neutrinos $E_\nu = 862 \text{ keV}$, the recast limit curve is analytically evaluated and depicted in Fig.~\ref{fig:borexino_limit}. In the light mass region where $M_{A'}^2 \ll 2m_e T_e$ ($M_{A'} \lesssim 100 \text{ keV}$), the propagator effects saturate ($\mathcal{F} \to 1$), and the recast limit degenerates exactly into the flat baseline value. Conversely, in the heavy mass regime where $M_{A'}^2 \gg 2m_e T_e$ ($M_{A'} \gtrsim 1 \text{ MeV}$), the shape factor analytically reduces to $\mathcal{F} \simeq 2m_e T_e / M_{A'}^2$. Consequently, the physical dipole cross-section is heavily suppressed by $(M_{A'})^{-4}$, causing the exclusion bound on $\mu_{\text{eff}}$ to relax quadratically as $M_{A'}$ increases---a kinematic characteristic distinctly captured by the steeply rising boundary in the figure.

While these kinematic characteristics objectively open up a parameter space for substantial dark magnetic moments to survive in the large mass region, this potential must be confronted with the stringent constraints on the visible-dark sector coupling. In our low-energy effective framework, the dark photon interacts with the visible sector exclusively via the kinetic mixing $\mathcal{L}_{\text{mix}} = - \frac{\epsilon}{2} F_{\mu\nu} F'^{\mu\nu}$. The magnitude of $\epsilon$ is tightly restricted by a synergistic combination of fixed-target and collider experiments.

For a sub-GeV $A'$, the dark photon predominantly decays invisibly into lighter dark states (e.g., $A' \to N_1 N_1$), producing a striking missing energy ($E_{\text{miss}}$) signature. The null observation of such events at the NA64 fixed-target experiment imposes exceptionally tight bounds, restricting $\epsilon \lesssim 10^{-5}$ for $m_{A'} \sim 10~\text{MeV}$ and $\epsilon \lesssim 10^{-4}$ near $m_{A'} \sim 100~\text{MeV}$~\cite{Banerjee:2019pds}. Once $m_{A'}$ surpasses the kinematic threshold of NA64, the mono-photon signature ($e^+e^- \to \gamma A'$) probed by BaBar provides the leading constraint, capping $\epsilon \lesssim 10^{-3}$ up to the $\Upsilon$ resonances~\cite{BaBar:2017tiz}.

\subsection{Joint Analysis and Cosmology}
\label{subsec:Joint Analysis}

The joint exclusion boundary on $\epsilon$ fundamentally creates a phenomenological tension. In the perturbative limit $\delta M_S \ll M_S$, the analytical prediction derived from our dual-scalar framework scales as:
\begin{equation}
    \mu_{ij} \simeq \frac{g_D Q_S^D m_N}{64\pi^2 M_S^2} \sin(2(\theta_I - \theta_R)) \cdot \text{Im}(\lambda_{j2}^* \lambda_{i1}^* - \lambda_{i2}^* \lambda_{j1}^*) \cdot \left( \frac{\delta M_S^2}{M_S^2} \right)^2 \cdot f\left( \frac{m_N^2}{M_S^2} \right) \mu_B.
\end{equation}
The theoretical upper bound on $\mu_{\text{eff}} = \epsilon \cdot \mu_{ij}$ thus experiences severe cross-constraints simultaneously from both sectors. Specifically, the MEG II limits on $\mu \to e \gamma$ strictly cap the flavor-violating Yukawa tensor $\text{Im}(\lambda_{j2}^* \lambda_{i1}^* - \lambda_{i2}^* \lambda_{j1}^*)$ to $\mathcal{O}(10^{-2})$ at the $M_E = 2\text{ TeV}$ benchmark. 

Fixing the kinematic benchmark at $M_S = 300\text{ MeV}$, $\delta M_S = 100\text{ MeV}$, and $m_N = 100\text{ MeV}$, an order-of-magnitude estimation sharply illustrates this dynamical conflict. Substituting the joint parameter bounds ($\epsilon \cdot y^2 \sim 10^{-6}$) and incorporating the fundamental loop suppression $1/(64\pi^2)$ alongside the kinematic factor $\mathcal{O}(10^{-3} \sim 10^{-4})$, the maximal achievable effective magnetic moment exhibits a strict single-parameter dependence on the dark gauge coupling:
\begin{equation}
    \mu_{\text{eff}}^{\text{max}} \sim \mathcal{O}(10^{-10}) \times g_D \cdot \mu_B.
\end{equation}

\begin{figure}[!htbp]
    \centering
    \includegraphics[width=0.95\textwidth]{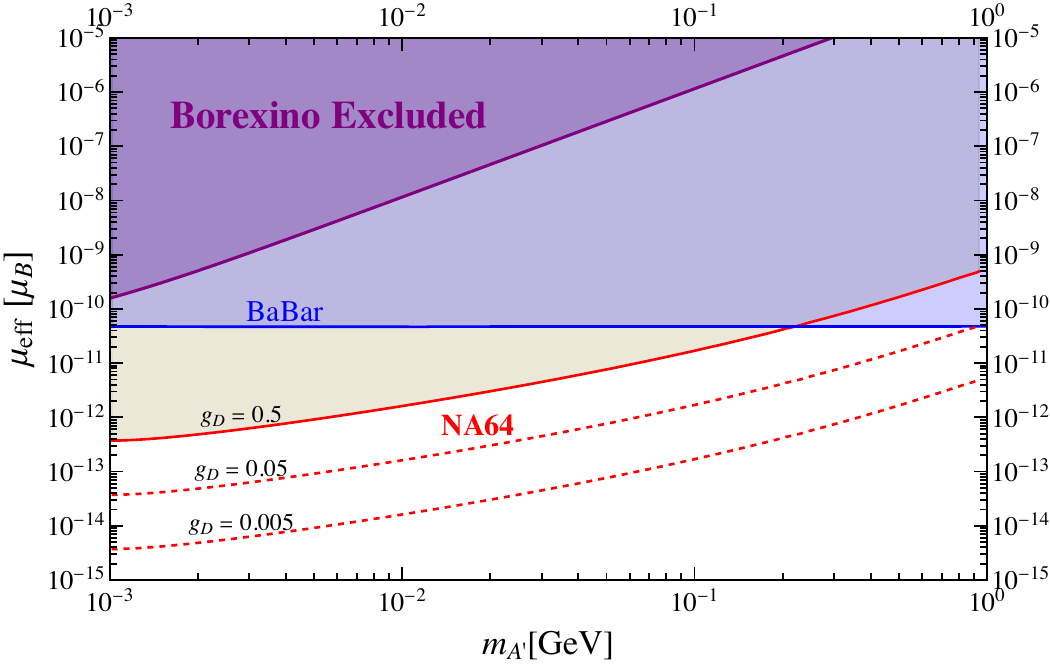}
    \caption{Joint exclusion plot of the model parameter space assuming benchmark masses $M_S = 300\text{ MeV}$, $\delta M_S = 100\text{ MeV}$, and $m_N = 100\text{ MeV}$. The figure compares the direct observational limits from Borexino~\cite{Agostini_2017} on the effective dark magnetic moment (purple solid line with shaded exclusion region) against the theoretical upper constraints jointly dictated by MEG II~\cite{megiicollaboration2024searchmutoegammadataset} and accelerator experiments. The blue lines depict the NA64~\cite{Banerjee:2019pds} + MEG II constraints evaluated at three distinct dark gauge couplings ($g_D = 0.5, 0.05, 0.005$), demonstrating the dominant exclusionary power in the sub-GeV mass window. The red solid line illustrates the complementary constraint from BaBar~\cite{BaBar:2017tiz} + MEG II at $g_D = 0.5$.}
    \label{fig:final_constraints}
\end{figure}

To rigorously quantify this interplay, we evaluate the complete one-loop analytical expression using the maximum parameters permitted by MEG II combined with the $\epsilon$ upper bounds from NA64 and BaBar. Fig.~\ref{fig:final_constraints} superimposes these theoretical indirect constraints directly against the Borexino exclusion boundary. We distinctly highlight the NA64 + MEG II joint constraint since NA64 provides the most stringent bounds in the low-mass region. 

The numerical results perfectly align with the analytical estimations. At a relatively large coupling of $g_D = 0.5$, the model's maximal prediction generated by the NA64 and BaBar limits marginally approaches the Borexino boundary. However, for smaller dark couplings ($g_D \lesssim 0.05$), the theoretical maximum dictated by NA64 + MEG II shifts downward entirely, falling far below the direct scattering constraints. This establishes a critical phenomenological hierarchy: for dark sector models featuring multi-generation scalar LFV mechanisms, indirect cLFV and accelerator searches jointly constitute the overwhelmingly dominant constraints. Any attempt to artificially amplify the neutrino magnetic moment signal to observable levels is fundamentally intercepted by these severe bounds, rendering the direct solar neutrino limits phenomenologically secondary within this microscopic framework.

In addition to the rigorous terrestrial probes discussed above, light dark sector particles are generally subject to astrophysical and cosmological bounds. However, for the benchmark mass regime of interest in this work ($M_{A'}, M_{S_a} \in [1~\text{MeV}, 1~\text{GeV}]$), stellar cooling constraints---derived from the Sun, red giants, and SN1987A~\cite{Raffelt:1996wa, Chang:2018rso, Capozzi:2020cbu}---naturally decouple. Even in core-collapse supernovae with core temperatures $T \sim 30~\text{MeV}$, the production of $\mathcal{O}(100)~\text{MeV}$ dark particles is severely kinematically forbidden by the Boltzmann suppression factor ($e^{-M/T}$).

Conversely, early-universe cosmology, specifically Big Bang Nucleosynthesis (BBN) and the effective number of relativistic species ($N_{\text{eff}}$), imposes strict boundaries on MeV-scale dark sectors~\cite{Fradette:2014sza, Planck:2018vyg, Cyburt:2015mya, Sabti:2020yrt, Depta:2020wmr}. If the dark and visible sectors ever thermalize, the late-time decay of heavy dark states into Standard Model particles can inject entropy into the electromagnetic bath, altering $N_{\text{eff}}$ and photodissociating light elements. Nevertheless, these cosmological bounds are highly sensitive to assumptions regarding the early-universe thermal history. Non-standard cosmologies, such as a low reheating temperature or scenarios where the dark sector temperature remains kinematically decoupled ($T_{\text{dark}} \neq T_{\text{SM}}$), can drastically relax these constraints. 

Furthermore, within our specific framework, the dominant invisible decay channels (e.g., $A' \to N_1 N_1$) efficiently confine the energy transfer entirely within the dark sector, significantly mitigating direct entropy injection into the visible plasma. Given these profound model dependencies, a rigorous evaluation of the cosmological phase diagram via coupled Boltzmann equations is deferred to future study. This work prioritizes the theoretical architecture of the misaligned double-scalar mechanism and its robust, less model-dependent validation through current terrestrial facilities like NA64, MEG II, and Borexino.

\section{Conclusions}
\label{sec:conclusion}
Driven by the vast disparity between the vanishingly small Standard Model predictions for Majorana neutrino magnetic moments and the sensitivities of ongoing direct detection experiments, this work proposed a novel dark sector framework to dynamically generate a macroscopic dark transition magnetic moment. Within the SM, the generation of a Majorana magnetic moment is heavily penalized by a GIM-like suppression and the necessity of an internal chirality flip proportional to the active neutrino masses. To systematically circumvent these structural bottlenecks, we extended the SM gauge symmetry with a secluded $U(1)_D$ group, introducing a Vector-Like Lepton doublet alongside two complex dark scalars. Unlike conventional scalar or vector portal models---where diagonal vector couplings strictly vanish for Majorana fermions and pure scalar couplings often face tight cosmological constraints---our framework elegantly realizes a tensor portal. The uniqueness of this architecture lies in its misaligned double-scalar mechanism. By intertwining the heavy vector-like fermions with two dark scalars exhibiting misaligned CP-even and CP-odd mixings, the required chirality flip is safely displaced onto the heavy internal fermion line, while the non-parallel flavor couplings dynamically break the Majorana antisymmetry prohibition.

To rigorously validate this theoretical construct, we performed a comprehensive phenomenological analysis confronting the model with a suite of multi-frontier experimental limits. A profound dynamical tension emerges within this framework because the macroscopic observable, the effective magnetic moment $\mu_{\text{eff}}$, is a direct product of the visible-dark kinetic mixing $\epsilon$ and the internal dark transition magnetic moment $\mu_{\text{tr}}$. We demonstrated that the generation of $\mu_{\text{tr}}$ is inextricably linked to charged lepton flavor violation in the visible sector, as both processes share the same underlying Yukawa couplings and heavy fermion mass scale. Consequently, the exceptionally stringent MEG II limits on the radiative decay $\mu \to e \gamma$ strictly cap the internal flavor-violating strength of the theory. Simultaneously, the kinetic mixing portal $\epsilon$ is severely bottlenecked by missing energy signatures at the NA64 fixed-target experiment in the sub-GeV regime and mono-photon searches at BaBar at higher masses. 

By analytically projecting these synergistic accelerator and cLFV constraints onto the parameter space of the effective magnetic moment, we established a definitive phenomenological hierarchy. Our numerical evaluations, anchored at typical sub-GeV scalar and heavy $\mathcal{O}(\text{TeV})$ fermion mass benchmarks, reveal that the theoretical upper bound on $\mu_{\text{eff}}$ is fundamentally intercepted by indirect limits before it can reach the detection threshold of solar neutrino experiments. Even when pushing the internal dark gauge coupling toward its perturbative limit, the maximal achievable magnetic moment remains heavily eclipsed by the direct scattering constraints from Borexino. Ultimately, this work highlights that for microscopic models relying on heavy fermions and multi-generation scalar LFV to enhance neutrino electromagnetic properties, high-intensity flavor probes and accelerator-based dark sector searches possess overwhelmingly dominant exclusionary power. Future scrutiny of such tensor portal paradigms will inevitably rely on the forthcoming generation of cLFV facilities, such as Mu3e and Mu2e, alongside a detailed mapping of their early-universe cosmological evolution.

\acknowledgments
We thank Jinhui Guo, Yan Luo, and Jia Liu for their insightful discussions during the early stages of this work.

\bibliographystyle{JHEP}
\bibliography{references}
\end{document}